\documentclass[a4paper,prd,nofootinbib,center]{revtex4}
\usepackage[dvips]{graphicx}
\usepackage{graphics}
\usepackage{dcolumn}
\usepackage{amssymb}
\usepackage{bm}
\usepackage{amsmath}
\usepackage{blindtext}
\usepackage{soul}
\usepackage{color}
\usepackage[dvipsnames]{xcolor}
\usepackage{subfigure}
\usepackage{epstopdf}
\usepackage{listings}
\usepackage{multirow}
\usepackage{ulem}
\usepackage{hyperref}
\hypersetup{
    colorlinks=true,
    linkcolor=blue,    
    urlcolor=magenta,
    citecolor=OliveGreen
}
\usepackage{cleveref} 
\usepackage[hang, flushmargin]{footmisc}
\usepackage{footnotebackref}

\begin{document}
 
\title{Structure formation and quasi-spherical collapse from initial curvature perturbations\\ with numerical relativity simulations}

\author{Robyn L.  Munoz${{^a}}$\footnote{robyn.munoz@port.ac.uk} \&
Marco Bruni${{^{a,b}}}$\footnote{marco.bruni@port.ac.uk} 
\vspace{0.5cm}}

\affiliation{${{^a}}$Institute of Cosmology {\rm \&} Gravitation, University of Portsmouth, Dennis Sciama Building, Burnaby Road, Portsmouth, PO1 3FX, United Kingdom\\
\\
${{^b}}$INFN Sezione di Trieste, Via Valerio 2, 34127 Trieste, Italy\\
}

\date{February 16, 2023 }

\begin{abstract}
We use numerical relativity simulations to describe the spacetime evolution during nonlinear structure formation in $\Lambda$CDM cosmology.
Fully nonlinear initial conditions are set at an initial redshift $z\approx 300$, based directly on the gauge invariant comoving curvature perturbation $\mathcal{R}_c$ commonly used to model early-universe fluctuations. 
Assigning a simple 3-D sinusoidal structure to $\mathcal{R}_c$, we then have a lattice of quasi-spherical over-densities representing idealised dark matter halos connected through filaments and surrounded by voids. 
This structure is implemented in the synchronous-comoving gauge, using a pressureless perfect fluid (dust) description of CDM, and then it is fully evolved with the Einstein Toolkit code. 
With this, we look into whether the Top-Hat spherical and homogeneous collapse model provides a good description of the collapse of over-densities.
We find that the Top-Hat is an excellent approximation for the evolution of peaks, where we observe that the shear is negligible and collapse takes place when the linear density contrast reaches the predicted critical value $\delta^{(1)}_C =1.69$. 
Additionally, we characterise the outward expansion of the turn-around boundary and show how it depends on the initial distribution of matter, finding that it is faster in denser directions, incorporating more and more matter in the infalling region. 
Using the EBWeyl code \cite{R.L.Munoz_M.Bruni_2022} we look at the distribution of the electric and magnetic parts of the Weyl tensor, finding that they are stronger along and around the filaments, respectively. 
We introduce a method to dynamically classify the different regions of the simulation box in Petrov types. 
With this, we find that the spacetime is of Petrov type I everywhere, as expected, but we can identify the leading order type in each region and at different times. 
Along the filaments, the leading Petrov type is D, while the centre of the over-densities remains conformally flat, type O, in line with the Top-Hat model. 
The surrounding region demonstrates a sort of peeling-off in action, with the spacetime transitioning between different Petrov types as non-linearity grows, with production of gravitational waves.
\end{abstract}

\maketitle

\section{Introduction} \label{sec: Introduction}
As small fluctuations in an otherwise homogeneous universe grow, they become the large-scale structures we observe today \cite{J.A.Peacock_1999, H.Mo_etal_2010, N.Vittorio_2018}. 
To describe this evolution non-linearly, multiple approaches have been created \cite{V.Sahni_P.Coles_1995, P.Monaco_1998, J.A.Peacock_1999, H.Mo_etal_2010, N.Vittorio_2018}, starting with the simple Top-Hat spherical and homogeneous collapse model \cite{J.E.Gunn_J.R.Gott_1972}. 
The Top-Hat describes a homogeneous spherical over-density in the matter-dominated era, with a dust fluid describing pressureless cold dark matter (CDM). 
This over-dense sphere is modelled by a closed (positive spatial curvature) FLRW ``separate universe" within an external FLRW background universe, usually spatially flat (zero curvature). 
The radius of the Top-Hat over-density expands at a slower rate than the background, gradually slowing down, as it is bound by its positive curvature (equivalent to the conserved and negative mechanical energy in the Newtonian description of the Top-Hat).
It eventually reaches its maximal size, turns around, and then contracts into itself to collapse. 
However simple this seems, the Top-Hat model provides the critical value of the linear density contrast corresponding to collapse, $\delta^{(1)}_{C}=1.69$, a crucial benchmark\footnote{ This value assumes that the cosmological constant $\Lambda$ is negligible, i.e.\ that the collapse occurs well before $\Lambda$ becomes relevant in the Friedman equation for the background. } to estimate virialisation and for the Press-Schechter mass function and the Sheth-Tormen extension \cite{W.H.Press_P.Schechter_1974, R.K.Sheth_G.Tormen_1999}.

More complex models have since been created with either inhomogeneity, a non-spherical shape, or with angular momentum \cite{H.Mo_etal_2010, V.Faraoni_2021}, most notably the Zel'dovich approximation, in the context of Newtonian structure formation, informs us on how pancakes are formed \cite{Y.B.Zel'dovich_1970} and how they represent the attractors for the dynamics \cite{M.Bruni_etal_2003_21Jan}. 
Yet, all these models lack relaxation mechanisms that would bring the structure to its final virialised stable state. 
The first attempt to describe this was with statistical mechanics \cite{D.Lynben-Bell_1967}, however, analytical limitations have led the field to work with numerical simulations instead. 
With these tools, an inhomogeneous universe is either modelled with a fluid or particle description of matter. 

N-body simulations have the advantage of going beyond shell crossing and inform us on the virialisation process and the shape of large-scale structures \cite{S.D.M.White_M.J.Rees_1978, J.F.Navarro_etal_1996, F.Pace_etal_2019, S.Saga_etal_2021, R.E.Angulo_O.Hahn_2022}.
With some caveat, it has been shown that in the Newtonian case the simulations accurately portray structure formation when compared with general relativistic simulations, except when the weak gravity regime doesn't hold \cite{W.E.East_etal_2018}. 
To make the description of gravity in N-body simulations somehow general relativistic, multiple approaches have been attempted.
A simple approximation has been used in \cite{G.Racz_etal_2017}, where matter is coupled to the expansion of distances with the average expansion-rate approximation.
A fully relativistic approach neglecting only tensor modes has been used in \cite{C.Barrera-Hinojosa_B.Li_2020_Jan, C.Barrera-Hinojosa_B.Li_2020_Apr, C.Barrera-Hinojosa_etal_2021_Jan}, based on the constant mean curvature and minimal distortion gauge.
In \cite{J.Adamek_etal_2016_Mar, J.Adamek_etal_2016_Jul} a weak field expansion has been used, based on the Poisson gauge with six degrees of freedom in the metric, see also \cite{J.Adamek_etal_2020}.
Alternatively, a relativistic post-processing treatment of Newtonian simulations can measure vector modes \cite{M.Bruni_etal_2014_Feb, D.B.Thomas_etal_2015_16Jul, C.Barrera-Hinojosa_etal_2021_Dec}, even for $f(R)$ gravity \cite{D.B.Thomas_etal_2015_30Jul}.
Finally, some relativistic effects can be extracted from Newtonian simulations with ray-tracing, see e.g.\ \cite{A.Barreira_etal_2016, Y.Rasera_etal_2021, C.Tian_etal_2022}.
To make the gravitational description fully relativistic, one may instead simplify the matter description and consider collisionless particles that evolve according to the global distribution \cite{C-M.Yoo_etal_2016, J.T.Giblin_etal_2019, W.E.East_etal_2019}. 
These types of simulations then meet similar challenges to fluid simulations.

The fluid description of matter lends itself more conveniently to the 3+1 formalism of numerical relativity \cite{W.E.East_etal_2012, J.M.Torres_etal_2014, J.Rekier_etal_2015, E.Bentivegna_M.Bruni_2016, J.T.Giblin_etal_2016, J.B.Mertens_etal_2016, H.J.Macpherson_etal_2017, J.Adamek_etal_2020}. 
While convenient for early times cosmology, together with scalar fields \cite{H.Kurki-Suonio_etal_1987, D.S.Goldwirth_T.Piran_1989, I.Musco_etal_2009, K.Clough_etal_2016,J.Braden_etal_2017, C-M.Yoo_etal_2019, J.C.Aurrekoetxea_etal_2020, T.Andrade_etal_2021, A.Ijjas_2022}, it finds its limitations at the first shell crossing. 
As structures decouple from the background and subsequently virialise, particles should go into a multi-stream regime, while in a fluid description shell crossing crashes simulations with comoving coordinates. 
Gauge choices can be made to avoid evolving such regions; however one main focus of this paper is on the collapse of over-densities and comparison with the Top-Hat model. 
Therefore, we work in the synchronous-comoving gauge, with the advantage of identifying collapse in terms of the proper time in the matter frame. 

The goal of this paper is to study the nonlinear evolution of the basic elements of the cosmic web, namely overdensities filaments \cite{J.R.Bond_etal_1995} and voids, extending the analysis in \cite{E.Bentivegna_M.Bruni_2016}, where a 3-dimensional (3-D) sinusoidal inhomogeneity in the matter density was evolved with varying amplitudes, and backreaction was found to be measurable, but extremely small. 
This periodic 3-D structure effectively represents a basic cosmic web, used also in \cite{E.Bentivegna_M.Bruni_2016, H.J.Macpherson_etal_2017, W.E.East_etal_2018, J.C.Aurrekoetxea_etal_2020, S.Saga_etal_2021}, a periodic lattice of over-densities (OD) and under-densities (UD), such that close to its peak each OD is approximately spherically symmetric. 
OD peaks are connected by over-dense filaments and are separated by voids, thus automatically satisfying the periodic boundary conditions that we use.
Here we evolve this 3-D structure in full General Relativity, describing CDM as a pressureless fluid with the same evolution codes in Einstein Toolkit \cite{F.Loffler_etal_2012, E.Bentivegna_2017, S.R.Brandt_etal_2020}.
However, we take a different approach to set the initial conditions, implementing the 3-D sinusoidal structure in the comoving curvature perturbation $\mathcal{R}_c$, originally introduced in \cite{D.H.Lyth_1985}. 
This is convenient because $\mathcal{R}_c$ is a gauge-invariant and time-independent variable at first order in perturbation theory and in the long wavelength approximation \cite{M.Bruni_etal_2014_Sep}, and it is commonly used to model inhomogeneities in the early universe, e.g.\ in inflationary models, see \cite{K.A.Malik_D.Wands_2008} and Refs.\ therein. 
Starting from the scalar potential $\mathcal{R}_c$, following the method described in \cite{M.Bruni_etal_2014_Mar} we set the initial spatial metric $\gamma_{ij}$ and the extrinsic curvature $K_{ij}$ as if these were first-order scalar perturbations, but then we treat them exactly, with no approximations, and use $\gamma_{ij}$ to compute the 3-Ricci scalar ${}^{(3)}R$ in full nonlinearity, and this ${}^{(3)}R$ and $K_{ij}$ are used in the Hamiltonian constraint to construct the matter density distribution $\rho$, so that the Hamiltonian constraint is automatically satisfied on the initial slice. 
By the same token, the momentum constraint is satisfied at first-perturbative order \cite{M.Bruni_etal_2014_Mar}.

This novel method to set up initial conditions for numerical relativity cosmological simulations has two advantages: {\it i)} it directly implements a purely growing mode, the only one that should exist in the early matter era and {\it ii)} it can be used to directly implement initial curvature perturbations predicted by inflationary models \cite{K.A.Malik_D.Wands_2008, M.Bruni_etal_2014_Mar, M.Bruni_etal_2014_Sep}. 
After summarising the necessary $\Lambda$CDM perturbations results \cite{M.Bruni_etal_2014_Mar} in Section~\ref{sec: Initial Distribution} our method of setting up nonlinear initial conditions and how they are implemented is described in Section~\ref{sec: fully nonlinear initial conditions}. 
 
Using this method we obtain a reliable evolution of the simple and reasonably realistic scenario provided by  the 3-D structure described above.
In particular for the non-spherical over-densities, whose evolution can be reliably compared to the Top-Hat model \cite{J.E.Gunn_J.R.Gott_1972, V.Sahni_P.Coles_1995, P.Monaco_1998, J.A.Peacock_1999, H.Mo_etal_2010, N.Vittorio_2018}. 
Our initial conditions depend on three parameters, namely the amplitude, wavelength, and initial redshift, whose impact on the initial inhomogeneities is explored in Section~\ref{sec: nonlin and long-wavelength regimes}.
The Fortran thorn ICPertFLRW \cite{R.L.Munoz_2023_ICPertFLRW} adapted to the Cactus code \cite{T.Goodale_etal_2003} was developed to implement these initial conditions in the Einstein Toolkit \cite{S.R.Brandt_etal_2020}; it is described in Section~\ref{sec: code description and numerical implementation}.

We describe the evolution at the centre of the OD and UD in Section~\ref{sec: Peak of the over-density} and, to explain this evolution, we consider the contributions to the Raychaudhuri equation in Section~\ref{sec: Raychaudhuri equation}.
We also look at how the turn-around (TA) boundary evolves, describing the infalling domain, in Section~\ref{sec: infalling domain}, and we consider the evolution of a domain contained within a comoving sphere of various comoving radii in Section~\ref{sec: Comoving sphere}.

Furthermore, our simulations are in full General Relativity, hence we also consider the gravitational description of our 3-D structure using the Weyl tensor. 
The electric and magnetic parts of the Weyl tensor \cite{A.Matte_1953, S.W.Hawking_1966, G.F.R.Ellis_2009, R.Owen_etal_2011, G.F.R.Ellis_etal_2012, R.Maartens_B.A.Bassett_1998, E.Bentivegna_etal_2018, A.Heinesen_H.J.Macpherson_2022} are computed in post-processing with EBWeyl, the code presented in Paper 1 \cite{R.L.Munoz_M.Bruni_2022, R.L.Munoz_2022_EBWeyl}, we then characterise the gravito-electromagnetic evolution of the 3-D structure in Section~\ref{sec: gravito-electromagnetism}. 
Additionally, the same code can be used to compute the invariants needed to classify the spacetime according to the Petrov type \cite{H.Stephani_etal_2003, P.Jordan_etal_1964}. The 3-D structure in our fully nonlinear simulations is general enough to find in Section~\ref{sec: Petrov classification} that the spacetime is of Petrov type I, as expected. 
We then introduce a novel method for the dynamical Petrov classification of different space regions by using thresholds: this enables us to define a leading-order Petrov type in each region and at different times.
in addition, we also show how this Petrov type depends on the shape of the inhomogeneity.

\textit{Assumptions \& notations}:  the speed of light is $c=1$,  the Einstein coupling constant is $\kappa=8\pi G$,  the Newton gravitational constant is $G=1$. 
Greek indices indicate spacetime $\{0,\;...\;3\}$ and Latin indices space $\{1,\;2,\;3\}$. 
Background quantities are given an overhead bar and the $(n)$ superscript is given to a perturbation of order $n$. Proper time derivatives are indicated with an overhead dot.

\section{Theoretical framework} \label{sec: Th framework}
In this paper, we will be using numerical relativity for cosmological simulations of the evolution of inhomogeneities in a $\Lambda$CDM universe, starting from initial data at a redshift $z_{IN}\sim 300$. 
In this section, we first summarise the fluid-flow description for the kinematics and dynamics of CDM, represented as a pressureless fluid (dust), and then we present the method that we use to set up initial conditions. 
Finally, we discuss how the initial amplitude and redshift of the inhomogeneities, together with the ratio of their length-scale to that of the Hubble scale, determine the change from linearity to non-linearity of the initial conditions, and the long-wavelength regime dominated by the spatial curvature perturbations.

\subsection{CDM as irrotational dust fluid}
In the 3+1 approach to numerical relativity \cite{M.Alcubierre_2008, T.W.Baumgarte_S.L.Shapiro_2010, M.Shibata_2015} the fundamental dynamical variables are the spatial metric $\gamma_{ij}$ and the extrinsic curvature $K_{ij}$, while  lapse $\alpha$ and shift $\beta^i$ represent the gauge freedom one has in propagating coordinates from one time slice to the next. 
In cosmology, a fundamental 4-vector field is always present, namely the 4-velocity of matter $u^\mu$, i.e.\ the eigenvector of the energy-momentum tensor\footnote{ This choice, called the energy frame, is not unique for imperfect fluids, see \cite{M.Bruni_etal_1992} and Refs.\ therein.} $T_{\mu\nu}$; here we will be dealing with pressureless CDM represented by $T_{\mu\nu}=\rho u^\mu u^\nu$, where $\rho$ is the rest-frame energy density of matter, a dust fluid.
In this paper, we will use the synchronous-comoving gauge such that $\alpha=1$, $\beta^i=0$ and $u^\mu=n^\mu$, where $n^\mu$ is normal to the time slices, so that $u^\mu = \{1,\;0,\;0,\;0\}$ and 
\begin{equation} \label{eq: metric}
    ds^2 = -d\tau^2 + \gamma_{ij} dx^i dx^j,
\end{equation}
where $\tau$ is the proper time, and in the following derivatives with respect to $\tau$ are denoted with an overhead dot.

In general, the kinematics of a fluid flow can be characterised by the variation $\nabla_\nu u_\mu$ of the 4-velocity $u^\mu$, defining kinematical quantities.
That is \cite{G.F.R.Ellis_H.van_Elst_1999, G.F.R.Ellis_2009, G.F.R.Ellis_etal_2012}, defining the projector $h_{\mu\nu} \equiv g_{\mu\nu} + u_\mu u_\nu$ orthogonal to $u^\mu$, we can decompose  $\nabla_\nu u_\mu$ in its irreducible parts
\begin{equation}
    \nabla_\nu u_\mu = \Theta_{\mu\nu} + \omega_{\mu\nu} - a_\mu u_\nu
    \;\;\;\;\;\;\;\;\;\;
    \Theta_{\mu\nu} = \frac{1}{3}h_{\mu\nu}\Theta + \sigma_{\mu\nu},
\end{equation}
where $a^{\mu} \equiv u^{\alpha} \nabla_{\alpha} u^{\mu}$ and  $\omega_{\mu\nu}\equiv  h^{\alpha}_{\mu}h^{\beta}_{\nu}\nabla_{[\beta} u_{\alpha]}$ are the 4-acceleration and the anti-symmetric vorticity tensor, and $\Theta_{\mu\nu}\equiv h^{\alpha}_{\mu}h^{\beta}_{\nu}\nabla_{(\beta} u_{\alpha)}$ is
the symmetric expansion tensor, decomposed into its trace and traceless parts, i.e.\ the expansion scalar $\Theta$ and the shear tensor $\sigma_{\mu\nu}$.
For dust, the 4-acceleration vanishes and fluid elements move along geodesics. 
In addition, with the choice of the synchronous-comoving gauge, the fluid is automatically irrotational ($\omega_{\mu\nu}=0$) and $h_{\mu\nu}$  and $\Theta_{\mu\nu}$ are purely spatial, with the first coinciding with $\gamma_{ij}$  and the second coinciding with the extrinsic curvature, so that $\Theta_{ij}=-K_{ij}=\frac{1}{2}\dot{\gamma}_{ij}$.
From its definition, for the expansion scalar $\Theta$ we can write 
\begin{equation} \label{eq: Theta def}
    \Theta \equiv \nabla_{\mu} u^{\mu} =  \frac{1}{\sqrt{\gamma}}\frac{\partial}{\partial x^\mu}\left(\sqrt{\gamma}u^\mu\right) = \frac{\dot{V}}{V},
\end{equation}
where the last equality holds in the synchronous-comoving gauge and $V=\sqrt{\gamma}$ is the local volume element, with $\gamma$ the determinant of the 3-metric $\gamma_{ij}$, so that in this gauge the expansion scalar coincides with the trace of the extrinsic curvature, $\Theta=-K=-K^i_i$.
Note that $a$

In general, from the conservation equations $\nabla_\mu T^{\mu\nu}=0$ one obtains the energy conservation and the momentum conservation equations projecting along and orthogonally to $u^\mu$, respectively. For dust, the momentum conservation is trivial and the energy conservation coincides with the continuity equation 
\begin{equation} \label{eq: continuity}
    \dot{\rho} = -\rho\Theta,
\end{equation}
where in general $\dot{\rho} = u^\alpha \nabla_\alpha \rho$, which in our gauge coincides with the partial derivative with respect to proper time.

Similarly defining $\dot{\Theta} = u^\alpha \nabla_\alpha \Theta$, the expansion scalar $\Theta$ satisfies the Raychaudhuri equation which, for irrotational dust, is
\begin{equation} \label{eq: Raychaudhuri}
    \dot{\Theta} = - \frac{1}{3}\Theta^2 - 2\sigma^2 - \frac{\kappa\rho}{2} + \Lambda,
\end{equation}
where $\sigma^2 = \sigma_{\mu\nu}\sigma^{\mu\nu}/2$, and $\Lambda$ is the cosmological constant.
Thus in general the Raychaudhuri and continuity equations are coupled to the evolution of the shear and of the electric and magnetic parts of the Weyl tensor \cite{G.F.R.Ellis_H.van_Elst_1999, G.F.R.Ellis_2009, G.F.R.Ellis_etal_2012}. 
Although we won't consider their evolution equations here, we will be dealing with their dynamics in Section~\ref{sec: gravito-electromagnetism}.

These quantities also satisfy various constraints \cite{G.F.R.Ellis_H.van_Elst_1999, G.F.R.Ellis_2009, G.F.R.Ellis_etal_2012}, here we only explicitly need the Hamiltonian constraint
\begin{equation} \label{eq: Hamiltonian}
    {}^{(3)}R + \frac{2}{3}\Theta^2 - 2\sigma^2 = 2\kappa\rho + 2\Lambda,
\end{equation}
where ${}^{(3)}R$ is the 3-Ricci scalar of the 3-metric $\gamma_{ij}$.

The continuity equation (\ref{eq: continuity}) just expresses conservation of the proper mass and, using Eq.~(\ref{eq: Theta def}), can be integrated to give 
\begin{equation}
    \rho\sqrt{\gamma} = \rho V = M(\mathbf{x})
\end{equation}
where $M(\mathbf{x})$ is the proper mass of the local fluid element. An integral of this quantity in a given coordinate domain will give the proper mass contained within that domain, see Appendix~\ref{sec: Num_int}.

\subsection{FLRW flat dust models}
In the case of a flat FLRW universe, we indicate quantities with an overhead bar: the spatial metric then is $\bar{\gamma}_{ij} = a^2 \delta_{ij}$ where $a=a(\tau)$ is the scale factor and $\delta_{ij}$ is the Kronecker delta, $H = \bar{\Theta}/3=\dot{a}/a$ is the Hubble expansion, $\bar{\rho} = 3H^2\Omega_{m}/\kappa$ is the energy density, where $\Omega_m$ is the dimensionless matter density parameter. 

Eq.~(\ref{eq: Raychaudhuri}) and Eq.~(\ref{eq: Hamiltonian}) reduce to the Friedmann equations and, together with Eq.~(\ref{eq: continuity}) these can be integrated in the flat $\Lambda$CDM case to get:
\begin{equation}\label{eq: LCDM}
    s = \left(\frac{\Omega_{m0}}{\Omega_{\Lambda 0}}\right)^{1/3}\sinh{\left(\frac{3 \tau H_0}{2}\sqrt{\Omega_{\Lambda 0}}\right)}^{2/3},
    \;\;\;\;\;\;\;\;\;\;\;\;
    H = H_0\sqrt{\Omega_{m0}s^{-3}+\Omega_{\Lambda 0}},
    \;\;\;\;\;\;\;\;\;\;\;\;
    \Omega_m = \Omega_{m0} / (\Omega_{m0} + \Omega_{\Lambda 0}s^3),
\end{equation}
where $H_0$ and $\Omega_{m0}$ are the values of these parameters today, and  $\Omega_{\Lambda 0} = \Lambda c^2 / 3 H_{0}^2 = 1 - \Omega_{m0}$ represents the cosmological constant contribution and $s=a/a_0$. In our simulations  we use the results from the Planck collaboration (2018) \cite{Planck_CMB_2018}: $\Omega_{m0}=0.3147$ and $cH_{0}^{-1} = 2997.9 \; h^{-1}$ Mpc, with $h=0.6737$.

We also consider the special case where $\Lambda=0$, i.e.\ the Einstein-de-Sitter model (EdS), where:
\begin{equation}\label{eq: EdS}
    s = \left(\frac{\tau}{\tau_0}\right)^{2/3},
    \;\;\;\;\;\;\;\;\;\;\;\;
    H = \frac{2}{3\tau},
    \;\;\;\;\;\;\;\;\;\;\;\;
    \Omega_m = 1.0.
\end{equation}

We emphasise that our simulations do not assume an overall $\Lambda$CDM or EdS expansion of the box domain, as in Newtonian N-body simulations, rather we use these models for comparison. 

\subsection{$\Lambda$CDM first-order perturbations} \label{sec: Initial Distribution}
Starting from \cite{J.M.Bardeen_etal_1983}, it is customary in the treatment of perturbations during inflation to introduce a variable that has the advantage of remaining constant while the perturbation scale is much larger than the Hubble scale, so that one can easily relate perturbations produced during inflation to when the same perturbations evolve in the radiation and matter eras, eventually re-entering the Hubble horizon. 
One such variable is the so-called ``gauge-invariant curvature perturbation on uniform density hypersurfaces" \cite{K.A.Malik_D.Wands_2008}
\begin{equation}\label{eq: zeta1}
    \zeta^{(1)} = -\mathcal{R}_c + \frac{1}{3}\delta^{(1)},
\end{equation}
where here $\delta^{(1)}$ represents the gauge-invariant first-order density perturbation\footnote{ The superscript $(1)$ denotes the perturbation order.} on comoving hypersurfaces, therefore automatically coinciding with the density contrast $\delta=\rho/\bar{\rho}-1$ in the synchronous-comoving gauge we use here, and $\mathcal{R}_c$ is the first-order gauge-invariant scalar perturbation potential for ${}^{(3)}R^{(1)}$, the first-order perturbation of the 3-Ricci scalar, see Eq.\ \eqref{eq: Rc_comoving_curvature_perturbation}. For reviews see \cite{K.A.Malik_D.Wands_2008} and \cite{D.Langlois_F.Vernizzi_2010}, where a fully nonlinear conserved quantity related to $ \zeta^{(1)}$ and $\mathcal{R}_c$ is also introduced.

In the following, we shall summarise the approach to perturbations in the synchronous-comoving gauge used in \cite{M.Bruni_etal_2014_Mar}, based on $\mathcal{R}_c$, in order to use this approach as  a starting point for our nonlinear initial condition set-up.  A parallel nonlinear long-wavelength approximation for inhomogeneities on large scales is used in \cite{M.Bruni_etal_2014_Sep}.
The advantage of using $\mathcal{R}_c$ as a starting point is twofold: {\it i)} it is directly related to  $\zeta^{(1)}$ by Eq.~\eqref{eq: zeta1} and it coincides with it at large scales, where $\delta^{(1)}$ is suppressed with respect to $\mathcal{R}_c$, see Eq.\ \eqref{eq: delta1} below; hence our set up for initial condition can be used to directly implement  perturbation predictions from inflationary models; {\it ii)} for dust, $\mathcal{R}_c$ is a conserved quantity at all times and for all scales, which can be used to implement all first-order scalar perturbations variables for the growing mode.
Let's consider scalar perturbations of a flat FLRW universe  in the matter-dominated era since these are the only relevant first-order perturbations for structure formation. 
In the synchronous-comoving gauge, and with Cartesian-like coordinates, the line element takes the form Eq.~(\ref{eq: metric}) and now we write the spatial metric $\gamma_{ij}$ as
\begin{equation} \label{eq: 3metric}
    \gamma_{ij} = a^2\left[(1-2\psi)\delta_{ij}+\chi_{ij}\right].
\end{equation}
The deviations from the FLRW background are $\psi$ and the trace-less $\chi_{ij}$, corresponding to the volume perturbation and anisotropic distortion respectively. Because we are only considering scalar perturbations, $\chi_{ij}$ at first-order is constructed from a scalar potential $\chi^{(1)}$ as follows:
\begin{equation}
    \chi_{ij} \simeq \left(\partial_{i}\partial_{j}-\frac{1}{3}\delta_{ij}\delta^{kl}\partial_{k}\partial_{l}\right)\chi^{(1)}.
\end{equation}

Then, $\psi$ is the only perturbation in the determinant of the spatial metric up to first order
\begin{equation}\label{eq: gdet}
    \gamma \simeq \bar{\gamma}(1-6\psi^{(1)}),
    \;\;\;\;\;\text{with}\;\;\;\;\;
    \bar{\gamma} = a^6.
\end{equation}
Given this metric, the first order perturbation to the 3-Ricci scalar, ${}^{(3)}R$, is associated with the comoving curvature perturbation $\mathcal{R}_c$ \cite{D.H.Lyth_1985} as
\begin{equation}\label{eq: Rc_comoving_curvature_perturbation}
    {}^{(3)}R^{(1)} = 4\nabla^2\mathcal{R}_c,
    \;\;\;\;\;\text{with}\;\;\;\;\;
    \mathcal{R}_c = \psi^{(1)}+\frac{a^2}{6}\nabla^2\chi^{(1)}.
\end{equation}
We remark that ${}^{(3)}R$ vanishes in any flat FLRW background, therefore according to the Stewart and Walker lemma \cite{J.M.Stewart_M.Walker_1974} cf.\ \cite{G.F.R.Ellis_M.Bruni_1989, M.Bruni_etal_1992, P.K.S.Dunsby_etal_1992}, ${}^{(3)}R^{(1)}$ and $\mathcal{R}_c$ are gauge-invariant, see Paper 1 \cite{R.L.Munoz_M.Bruni_2022} for a general discussion on invariant quantities.
The Laplacian $\nabla^2 = \gamma^{ij}\nabla_i\nabla_j$ is such that for first-order scalar perturbations, it takes the form $a^{-2}\delta^{ij}\partial_i\partial_j$. 
It can be shown that $\mathcal{R}_c$ is constant in time \cite{M.Bruni_etal_2014_Mar}, so that ${}^{(3)}R^{(1)} \propto a^{-2}$. 
Then, the starting point to express the first order perturbations $\delta^{(1)}$, $\psi^{(1)}$ and $\chi^{(1)}$ as a function of $\mathcal{R}_c$, is to consider \cite{M.Bruni_etal_2014_Mar} the evolution of the density contrast
\begin{equation}\label{eq: delta_evo_eq}
    4H\dot{\delta}^{(1)}+6H^2\Omega_m\delta^{(1)}={}^{(3)}R^{(1)},
\end{equation}
which can be derived from the continuity equation (\ref{eq: continuity}) and the Hamiltonian constraint Eq.~(\ref{eq: Hamiltonian}). 
Eq.~(\ref{eq: delta_evo_eq}) has two solutions: the homogeneous one, corresponding to the Hubble expansion, $\delta_- \propto H$, and therefore called the decaying mode, and the particular solution, the so-called growing mode $\delta_+$ sourced by the 3-curvature, and as such related to $\mathcal{R}_c$. 
By solely considering the growing mode Eq.~(\ref{eq: delta_evo_eq}) can be rearranged by  introducing the growth factor $f_1=d\ln\delta/d\ln a \simeq \Omega_{m}^{6/11}$ \cite{P.J.E.Peebles_1980, L.Wang_P.J.Steinhardt_1998}, to express $\delta^{(1)}$ as a function of $\mathcal{R}_c$
\begin{equation}\label{eq: delta1}
    \delta^{(1)} = \frac{\nabla^2\mathcal{R}_c}{FH^2},
\end{equation}
with $F = f_1+\frac{3}{2}\Omega_m$; in the early-matter era, when the EdS model is a good approximation and $\Omega_m = 1$, $f_1 = 1$ and $\delta^{(1)} \propto a$. 
With Eq.~(\ref{eq: delta1}), $\psi^{(1)}$ and $\chi^{(1)}$ can be expressed by using the deformation $\vartheta^{(1)}$. 
The expansion tensor, $\Theta_{ij}$, has a background part $\bar{\Theta}_{ij} = a^2 H \delta_{ij}$ and a perturbed part, the deformation tensor $\vartheta_{ij}$, such that $\Theta_{ij}=\bar{\Theta}_{ij}+\vartheta_{ij}$, with the trace $\Theta = \bar{\Theta} + \vartheta$, where $\bar{\Theta} = \bar{\gamma}^{ij}\bar{\Theta}_{ij} = 3H$. 
Additionally, in the synchronous-comoving gauge, the expansion tensor can be expressed as $\Theta_{ij}=\frac{1}{2}\dot{\gamma}_{ij}$, then the first order trace is $\vartheta^{(1)} = -3\dot{\psi}^{(1)}$. 
Likewise, the first order continuity equation is $\dot{\delta}^{(1)}=-\vartheta^{(1)}$. 
Then, putting these two expressions together $\dot{\delta}^{(1)}=3\dot{\psi}^{(1)}$, and so $\psi^{(1)}$ can be expressed as a function of $\mathcal{R}_c$ using Eq.~(\ref{eq: delta1}), where the integration constant is identified to be $\mathcal{R}_c$ from Eq.~(\ref{eq: Rc_comoving_curvature_perturbation}). 
Furthermore, $\psi^{(1)}$ can be introduced into Eq.~(\ref{eq: Rc_comoving_curvature_perturbation}) to provide $\chi^{(1)}$, such that
\begin{equation}\label{eq: psi1chi1}
    \psi^{(1)} = \frac{1}{3}\delta^{(1)}+\mathcal{R}_c,
    \;\;\;\;\;\text{and}\;\;\;\;\;
    \chi^{(1)} = -\frac{2 \mathcal{R}_c}{a^2 FH^2}.
\end{equation}
Therefore the spatial metric perturbed with a purely growing mode expressed up to first order as a function of $\mathcal{R}_c$ is
\begin{equation}\label{eq: metricRc}
    \gamma_{ij} = \bar{\gamma}_{ij} + \gamma_{ij}^{(1)} = a^2(1-2\mathcal{R}_c)\delta_{ij}-\frac{2}{FH^2}\partial_i\partial_j\mathcal{R}_c.
\end{equation}
With our synchronous-comoving gauge choice, the extrinsic curvature $K_{ij}=-\Theta_{ij}=-\frac{1}{2}\dot{\gamma}_{ij}$. 
Introducing Eq.~(\ref{eq: delta1}) into Eq.~(\ref{eq: delta_evo_eq}) shows that $\frac{d}{d\tau}(1/FH^2) = (2+f_1)/FH$ and since $\mathcal{R}_c$ is time independent
\begin{equation}\label{eq: curvRc}
    K_{ij} = \bar{K}_{ij} + K_{ij}^{(1)} = -a^2H(1-2\mathcal{R}_c)\delta_{ij}+\frac{(2+f_1)}{FH}\partial_i\partial_j\mathcal{R}_c.
\end{equation}
$K_{ij}$ can be separated into its trace $K$ and traceless $A_{ij}$ part
\begin{equation}
    K_{ij}=A_{ij}+\frac{1}{3}\gamma_{ij}K,
\end{equation}
such that in this gauge both are related to the fluid kinematical quantities. 
$A_{ij}$ is associated to the shear tensor of the matter flow $\sigma_{ij}$, $A_{ij}=-\sigma_{ij}$, at first order
\begin{equation}
    A_{ij}^{(1)} = -\sigma_{ij}^{(1)} = \frac{f_1}{FH}\left(\partial_i\partial_j - \frac{1}{3}\delta_{ij}\delta^{kl}\partial_k\partial_l\right)\mathcal{R}_c.
\end{equation}
We remark, that in the background $\bar{\sigma}_{ij}=0$, hence the shear is a first-order gauge invariant quantity.
Then, $K$ is associated to the expansion scalar $\Theta$:
\begin{equation}\label{eq: K}
    K = -\Theta = \bar{K}-\vartheta,
    \;\;\;\;\;\text{with}\;\;\;\;\;
    \bar{K} = -3H,
    \;\;\;\;\;\text{and}\;\;\;\;\;
    K^{(1)} = -\vartheta^{(1)} = f_1 H \delta^{(1)}.
\end{equation}
In this gauge the momentum density $J^i=0$, this means that the momentum constraint takes the form $D_{i}(K^{i}_{j})-D_{j}(K)=0$. 
It was shown \cite{M.Bruni_etal_2014_Mar} that at first order this expression reduces to $D_{j}(\dot{\mathcal{R}}_c)=0$, and since for dust $\dot{\mathcal{R}}_c=0$ at all times at all scales at first order, then at this order the momentum constraint is automatically satisfied.

As $\delta\equiv\rho/\bar{\rho}-1$ is the density contrast for the matter field, we can define similar quantities for the contrast of the volume element $\gamma$ and expansion $K$:
\begin{equation} \label{eq: pert}
    \delta \gamma \equiv \gamma / \bar{\gamma} - 1,
    \;\;\;\;\;\text{and}\;\;\;\;\;
    \delta K \equiv K / \bar{K} - 1.
\end{equation}
Given Eq.~(\ref{eq: gdet}), Eq.~(\ref{eq: psi1chi1}), and Eq.~(\ref{eq: K}) these can be expressed at first order as:
\begin{equation} \label{eq: delta_gamma1_K1}
    \delta \gamma^{(1)}=-6\left(\frac{1}{3}\delta^{(1)}+\mathcal{R}_c\right)
    \;\;\;\;\;\text{and}\;\;\;\;\;
    \delta K^{(1)}=-\frac{f_1\delta^{(1)}}{3}.
\end{equation}

\section{Fully nonlinear initial conditions} \label{sec: fully nonlinear initial conditions}
\subsection{Motivations}
In the standard scenario for the generation of structure formation in cosmology, the seeds are produced at large scales, well outside the Hubble horizon, during the inflationary epoch; these scales then re-enter the horizon  when the accelerated phase ceases and the seeds can grow.
More precisely, inflation produces an almost scale-invariant spectrum of  fluctuations in the metric variable $\zeta$, with the line element written as 
\begin{equation}\label{gammatilde}
    ds^2 = -d\tau^2 + a^2(\tau) e^{2\zeta(\tau,\;x^i)} \widetilde{\gamma}_{ij} dx^i dx^j,
\end{equation}
where $\text{det}(\widetilde{\gamma}_{ij})=1$, see \cite{K.A.Malik_D.Wands_2008, D.Langlois_F.Vernizzi_2010, M.Bruni_etal_2014_Mar, M.Bruni_etal_2014_Sep} and Refs.\ therein. 
In this scenario, $\zeta$ is nonlinear, but coincides with $\zeta^{(1)}$ in Eq. \eqref{eq: zeta1}. 
At large scales, in the long-wavelength approximation (AKA gradient expansion), at leading order $\zeta$ is constant and $ \widetilde{\gamma}_{kj} \simeq \delta_{kj}$, so that in this approximation the spatial metric in Eq.\ \eqref{gammatilde} is conformally flat, and the 3-Ricci scalar is then given by a beautifully simple expression in terms of $\zeta$ and its gradients \cite{M.Bruni_etal_2014_Sep}; at first perturbative order this expression simplifies to Eq.\ \eqref{eq: Rc_comoving_curvature_perturbation} above, and $\zeta^{(1)}=\mathcal{R}_c$ at large scales, where $\delta^{(1)}$ is suppressed in Eq.\ \eqref{eq: zeta1}. 
It actually turns out \cite{M.Bruni_etal_2014_Sep} that at leading order in this large-scales approximation, the equations for the inhomogeneities are formally exactly the same as those for first-order perturbations \cite{M.Bruni_etal_2014_Mar}. 
This nonlinear $\zeta$ is also used to model the birth of primordial black holes, see \cite{M.Shibata_M.Sasaki_1999, M.Ilia_2019} and Refs.\ therein, c.f.\ \cite{K.Clough_etal_2016, J.C.Aurrekoetxea_etal_2022, M.Corman_W.E.East_2022} for different approaches in numerical relativity.
In single-field slow-roll inflation, the primordial $\zeta$ is an almost Gaussian random field \cite{J.Maldacena_2003, V.Acquaviva_etal_2003}. 
In practice, therefore, non-Gaussianities are commonly modelled in terms of an expansion of $\zeta$ in terms of $\zeta^{(1)}$, parameterised by $f_{NL}$ and higher order parameters, $\zeta = \zeta^{(1)} + f_{NL}\zeta^{(1)2} + \dotsi$. 
Motivated by these standard modelling of primordial inhomogeneities,  we now set up fully nonlinear initial conditions using the scalar curvature variable $\mathcal{R}_c$. 

\subsection{Ansatz and implementation}
To this end, to set up initial conditions we have developed a new thorn ICPertFLRW \cite{R.L.Munoz_2023_ICPertFLRW}. 
The starting ansatz is that the metric and the extrinsic curvature are precisely given by their expressions Eq.~(\ref{eq: metricRc}) and Eq.~(\ref{eq: curvRc}), but should otherwise be thought of as quantities to be used in full non-linearity, generated by the scalar  potential $\mathcal{R}_c$.
From $\gamma_{ij}$ and $K_{ij}$, we then compute the 3-Ricci scalar ${}^{(3)}R$, the trace $K$, and the magnitude $K^{ij}K_{ij}$.
Given our ansatz, based on $\mathcal{R}_c$ and its derivatives, these quantities are computed analytically by ICPertFLRW \cite{R.L.Munoz_2023_ICPertFLRW}.
We can then use the Hamiltonian constraint to compute the initial matter density 
\begin{equation}\label{eq: rhoHam}
    \rho = \frac{1}{2\kappa}\left({}^{(3)}R + K^2 - K^{ij}K_{ji} - 2\Lambda \right) = \frac{1}{2\kappa}\left({}^{(3)}R + \frac{2}{3}K^2 - 2A^2 - 2\Lambda \right),
\end{equation}
with $A^2=A^{ij}A_{ji}/2$. 
We emphasise that in setting up initial conditions in full non-linearity, we introduce vector and tensor modes, in particular in the shear $\sigma_{ij}=-A_{ij}$ that sources the magnetic part of the Weyl tensor $B_{ij}$: this is non-zero, as it will be shown in Section~\ref{sec: gravito-electromagnetism}, while at first order $B_{ij}^{(1)}=0$ (in all gauges) for the purely scalar perturbation of the previous section.

The main advantage of using the Hamiltonian constraint to set up the initial distribution of the matter density $\rho$ in Eq.~(\ref{eq: rhoHam}) is twofold: 
\textit{i)} its algebraic use makes the constraint automatically satisfied in the initial time step,
\textit{ii)} in order to set up the initial conditions we don't need to solve an elliptic equation, as it is the case if the starting point is the distribution of $\rho$ itself, as in \cite{E.Bentivegna_M.Bruni_2016}. 
The Hamiltonian constraint was also used to non-linearly provide $\rho$ in \cite{J.T.Giblin_etal_2016}, although not using $\mathcal{R}_c$.
Note that we could have set up initial conditions exclusively using first-order quantities: we emphasise the benefit of our fully nonlinear method in Appendix~\ref{sec: Constraints errors convergence}, where we show that even starting from small initial perturbations nonlinear effects are important in General Relativity.

\begin{figure}[th]
    \centering
    \begin{minipage}{.49\textwidth}
        \centering
        \includegraphics[width=\linewidth]{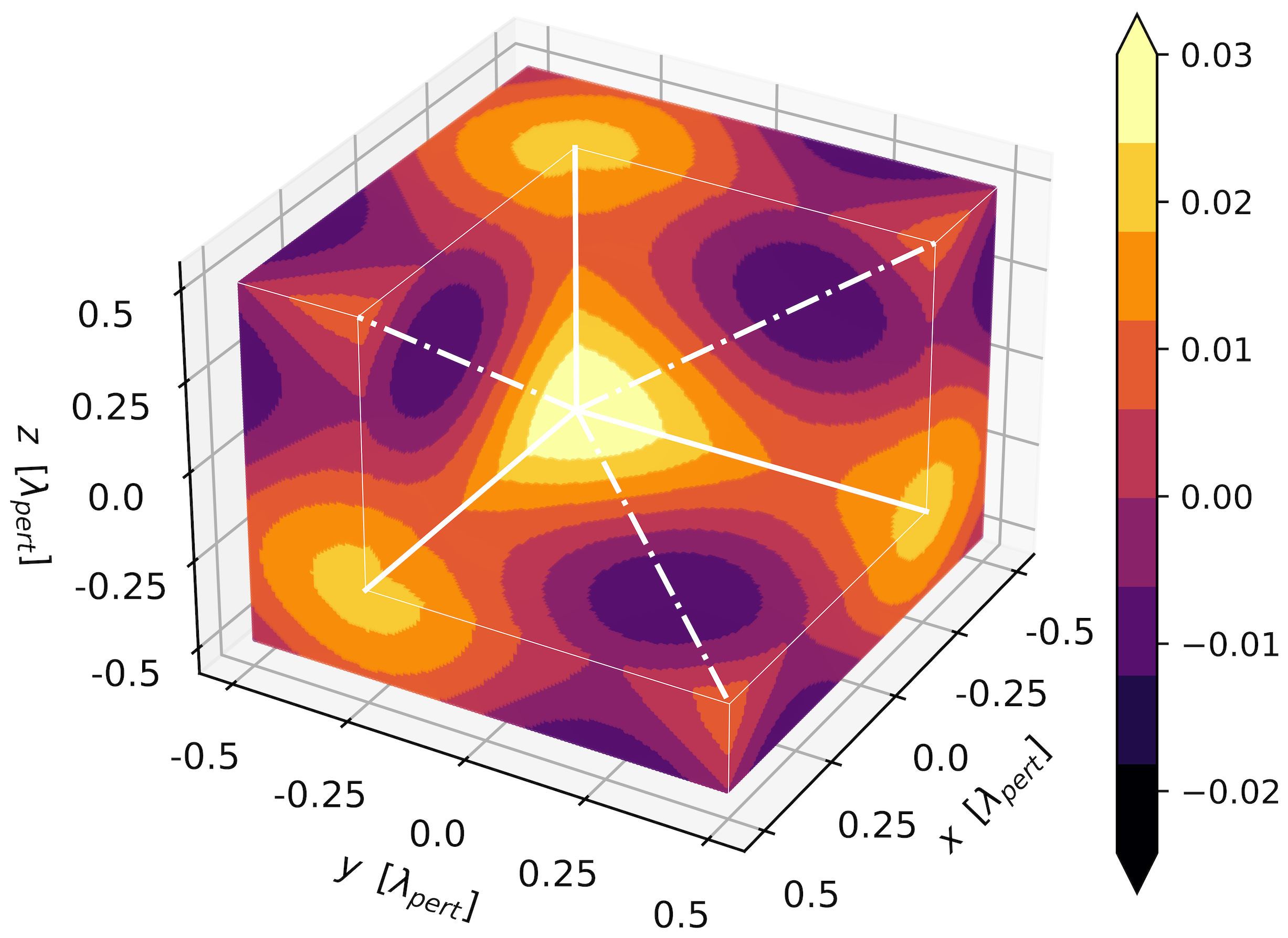}
        \caption{Initial distribution at $z_{IN} = 302.5$ of the density contrast $\delta$ in the simulation box, for a $\Lambda$CDM universe. The $x$, $y$, and $z >-0.25\lambda_{pert}$ region is removed exposing the centre of the over-density at $x=y=z=-0.25\lambda_{pert}$, where $\delta_{IN,\;OD}=0.03$. The full lines go through the vertices and dash-dotted lines through the centre of the edges of an octahedron centred at the over-density.}
        \label{fig: IniDelta3d}
    \end{minipage}
    \hspace{0.15cm}
    \begin{minipage}{.49\textwidth}
        \centering
        \includegraphics[width=\linewidth]{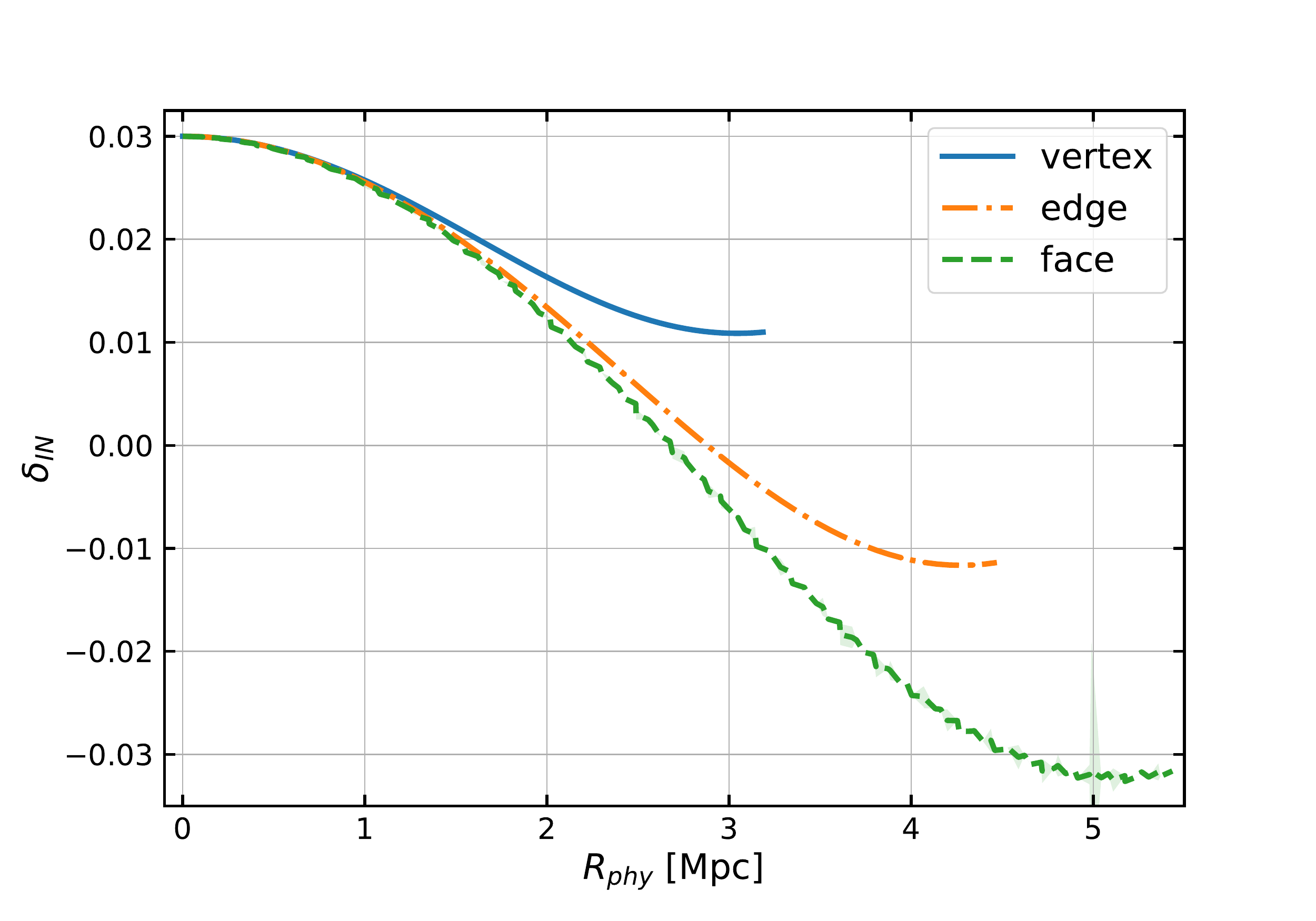}
        \caption{Initial radial profile at $z_{IN} = 302.5$ of the initial density contrast $\delta$ starting from the centre of the over-density to its minimum in three different directions, towards the vertices, edges, and faces of the octahedral distribution in Eq.~(\ref{eq: Rc}) plotted against the proper radius from the over-dense peak. Error bars, when visible, are indicated as shaded regions.}
        \label{fig: IniDelta1d}
    \end{minipage}
    \begin{minipage}{\linewidth}
        \centering
        \includegraphics[width=\linewidth]{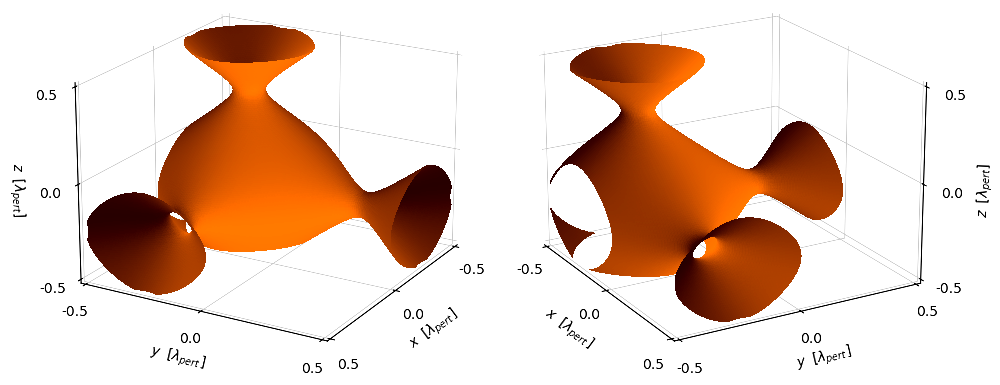}
        \caption{Isosurface for $\delta=0.01$ in the initial distribution of the matter density contrast at $z_{IN} = 302.5$. The two different panels show different points of view. The periodic boundary conditions insure that this distribution is a lattice of over-densities connected by filaments and separated by voids.}
        \label{fig: Isocurve}
    \end{minipage}
\end{figure}

All that remains is to define the comoving curvature perturbation $\mathcal{R}_c$.
A fully realistic initial set-up should consist of generating a spatial realisation of $\mathcal{R}_c$ starting from a Gaussian (or quasi-Gaussian) scale-invariant spectrum, but this is beyond our current scopes. 
Instead, we chose a single 3-D sinusoidal mode:
\begin{equation} \label{eq: Rc}
    \mathcal{R}_c = A_{pert}\bigg(\sin\left(x k_{pert}\right) + \sin\left(y k_{pert}\right) + \sin\left(z k_{pert}\right) \bigg),
\end{equation}
with $k_{pert} = 2\pi/\lambda_{pert}$ and the simulation box spanning $x, \; y, \; z\in[-\lambda_{pert}/2, \; \lambda_{pert}/2]$. 
$\lambda_{pert}$ is the comoving wavelength at the reference redshift $a(z_R)=1$, such that the physical wavelength is retrieved as $\lambda_{phy}=a \lambda_{pert}$. 
We work with $a(z_R=0)=1$ so that the comoving wavelength corresponds to a physical wavelength today, as defined in a reference $\Lambda$CDM FLRW spacetime, which would be the background in a perturbative setting.

A simulation box containing a ``compensated inhomogeneity", i.e.\ one as that in Eq.~\eqref{eq: Rc}, such that its linear average vanishes, essentially expands as the reference FLRW spacetime, i.e.\ backreaction is negligibly small \cite{E.Bentivegna_M.Bruni_2016, J.T.Giblin_etal_2019, H.J.Macpherson_etal_2019, J.Adamek_etal_2018}.
However, we emphasise that in general, averaged quantities do not exactly coincide with those of the FLRW model: even in the initial conditions, the non-linearity of General Relativity implies that the nonlinear average of Eq.~\eqref{eq: Rc} is non-zero. 
Furthermore, if a spatial region of a given comoving scale contains an OD that grows non-linearly, then its physical size today\footnote{ The size agreed by a network of comoving observers with synchronised clocks. } will eventually be much smaller than the corresponding FLRW physical scale.

The spatial distribution Eq.~\eqref{eq: Rc} allows us to focus on some specific relativistic features that emerge clearly in this simple set-up, features that would be probably harder to characterise in a more realistic scenario. 
Specifically, it will enable us to study the growth of an OD whose centre is at $x=y=z=-\lambda_{pert}/4$ and an UD whose centre is at $x=y=z=\lambda_{pert}/4$. It produces the initial $\delta$ presented in Fig.~(\ref{fig: IniDelta3d}, \ref{fig: IniDelta1d}, \ref{fig: Isocurve}). 
Fig.~(\ref{fig: IniDelta3d}) shows the initial $\delta$ distribution in the simulation box with the centre of the OD exposed, while Fig.~(\ref{fig: Isocurve}) shows the isosurface where $\delta=0.01$. These figures emphasise the non-spherical shape of this distribution. 
Indeed, the equation $\sum_{i=1}^{3}\sin(x^i k_{pert})=1$ parameterises an octahedron, so when close to the peak of the OD, spherical symmetry is approximated, further out an octahedron geometry creates filamentary-like structures periodically connecting each OD peak. 
We satisfy the boundary conditions by using periodic boundaries. However, we emphasise that the non-spherical nature of the distribution is not due to the boundary conditions in the simulation \cite{G.Racz_etal_2021}, but due to the choice of the initial distribution. 
 
Centring an octahedron around the OD we identify three main directions of interest from the centre of the OD: along the vertices, the centre of the edges and the centre of the faces. 
A half period of $\delta$ along each direction is presented in Fig.~(\ref{fig: IniDelta1d}). 
Close to the peak of the OD, the three directions overlap, highlighting the proximity to spherical symmetry. 
Beyond that, we see the axis going through the vertices never goes through an UD region, since this direction goes through the filaments (full white lines in Fig.~(\ref{fig: IniDelta3d}), and full blue lines in Fig.~(\ref{fig: IniDelta1d})), and the axis going through the centre of the faces goes through the centre of the UD (not in Fig.~(\ref{fig: IniDelta3d}), and green dashed lines in Fig.~(\ref{fig: IniDelta1d})). 
Although the spatial distribution that we derive from Eq.~(\ref{eq: Rc}) is unrealistic, it contains the three basic elements of the cosmic web, namely ODs, filaments \cite{J.R.Bond_etal_1995}, and voids and as such can be viewed as a skeleton description of large-scale structures and it is more realistic than the spherical Top-Hat model.

\subsection{Nonlinear and long-wavelength regimes} \label{sec: nonlin and long-wavelength regimes}
\begin{figure}[t]
    \centering
    \includegraphics[width=\linewidth]{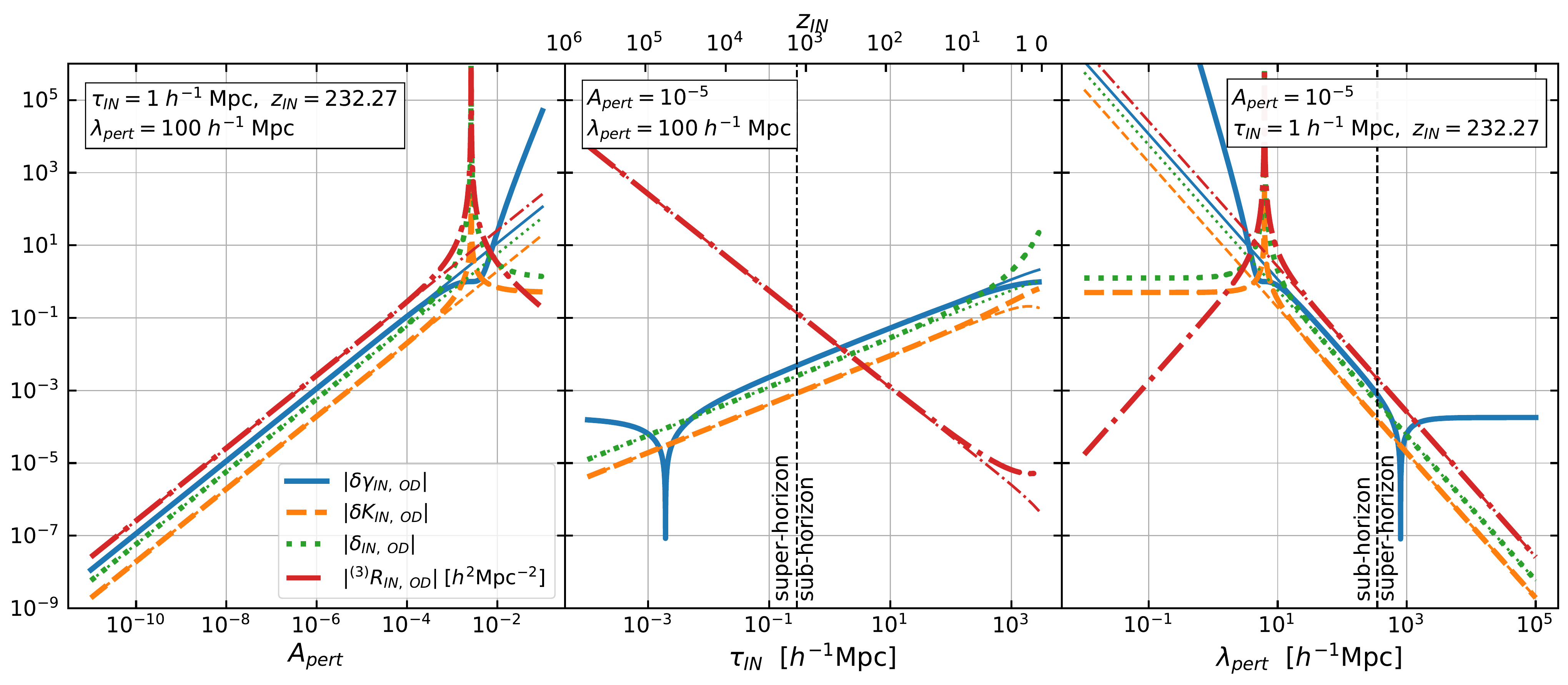}
    \caption{Amplitude of initial (IN) $\delta\gamma$, $\delta K$, $\delta$ and ${}^{(3)}R$ in the centre of the over-density (OD) as a function of $A_{pert}$, $z_{IN}$ and $\lambda_{pert}$ presented in each panel left to right. 
    While each is varied the other parameters are kept constant as presented in the top box. 
    The thinner lines correspond to the first-order expressions of these quantities, while the thicker lines correspond to their nonlinear expressions, thus the separation of these two lines emphasises non-linearity. 
    The vertical dashed black lines indicate the instance where the physical wavelength corresponds to the Hubble distance $\lambda_{phy}=c/H$ hence separating sub and super Hubble horizon regimes. 
    \textit{Left panel:} for the given initial redshift $z_{IN}$ and perturbation wavelength $\lambda_{pert}$, non-linearities start to be relevant when $A_{pert}>10^{-4}$. 
    \textit{Middle panel:} for the given $A_{pert}$ and $\lambda_{pert}$ non-linearities would only be relevant for $z_{IN} \lesssim 50$. 
    The first-order thin lines become curved when $\Lambda$ becomes relevant. The proper volume perturbation $\delta\gamma_{IN,\;OD}$ shows a plateau during the $\mathcal{R}_c$-dominated regime, see Eq.~(\ref{eq: delta_gamma1_K1}) and Eq.~(\ref{eq: Long_wavelength}), when $\delta\gamma_{IN,\;OD}>0$, and its sign changes in the transition to the $\delta$-dominated regime $\delta\gamma_{IN,\;OD}<0$.
    \textit{Right panel:} for the given $A_{pert}$ and $z_{IN}$ non-linearities are only relevant on scales smaller than $\lambda_{pert}\lesssim\text{few}\times 10 \text{h}^{-1}\text{Mpc}$. The $\mathcal{R}_c$-dominated regime is again identifiable with the plateau in $\delta\gamma_{IN,\;OD}$ on large scales.
    }
    \label{fig: PertAmp}
\end{figure}

The above initial distribution lets us freely choose the amplitude and wavelength of the inhomogeneity, $A_{pert}$ and $\lambda_{pert}$, as well as the initial redshift $z_{IN}$. 
The impact of these parameters on the initial amplitude of $\delta \gamma$, $\delta K$, $\delta$ and ${}^{(3)}R$ at the peak of the OD is presented in Fig.~(\ref{fig: PertAmp}). 
The thin lines are the first-order quantities from Eq.~(\ref{eq: Rc_comoving_curvature_perturbation}), Eq.~(\ref{eq: delta1}) and Eq.~(\ref{eq: delta_gamma1_K1}) whereas the thick lines are the fully nonlinear quantities obtained from Eq.~(\ref{eq: metricRc}), Eq.~(\ref{eq: curvRc}), Eq.~\eqref{eq: pert} and Eq.~(\ref{eq: rhoHam}). 
Each panel shows their dependencies on $A_{pert}$, $z_{IN}$ and $\lambda_{pert}$ (left to right respectively) while keeping the other two parameters constant (with their values listed in the top box).

In the left panel, we consider inhomogeneities on a scale well inside the Hubble horizon at that time. 
This shows that the inhomogeneities are proportional to $A_{pert}$ when $A_{pert}$ is small enough. 
However, when $A_{pert}$ is large there is a separation between the thick and thin lines: this identifies the emergence of the nonlinear regime. 
This is also visible in the other panels for low redshift and small scales, domains where local dynamics become dominant.
Otherwise, inhomogeneities in the linear regime are given by the Laplacian of $\mathcal{R}_c$ and as such, they are proportional to $\lambda_{pert}^{-2}$ for the right panel and proportional to $a(\tau)$ in the middle panel, except ${}^{(3)}R \propto a^{-2}(\tau)$.
In the middle panel, at low redshift linear curves are no longer straight because in $\Lambda$CDM we depart from the $\delta$-dominated era.

We emphasise that the inhomogeneity in the proper volume at the OD $\delta \gamma_{IN,\;OD}$ has a peculiar dependence on $A_{pert}$, $z_{IN}$ and $\lambda_{pert}$ even in the linear regime, as clearly visible in the middle and right panels in Fig.~(\ref{fig: PertAmp}).
To understand this, consider Eq.~(\ref{eq: delta_gamma1_K1}), which shows that $\delta \gamma^{(1)}$ is composed of two terms: $\mathcal{R}_c$ and $\delta^{(1)}$. 
Given the $\mathcal{R}_c$ sinusoidal distribution Eq.~(\ref{eq: Rc}), the Laplacian in $\delta^{(1)}$, Eq.~(\ref{eq: delta1}), creates a sign difference between these two terms. 
$\delta \gamma$ then has $\mathcal{R}_c$-dominated and $\delta$-dominated regimes and the transition is highlighted by a sign change (the downward spike in the log-plot Fig.~(\ref{fig: PertAmp})). 
$\mathcal{R}_c$ and $\delta^{(1)}$ are both proportional to $A_{pert}$, which can even be factored out in Eq.~(\ref{eq: delta_gamma1_K1}), so that the relative weight of $\mathcal{R}_c$ and $\delta^{(1)}$ in the left panel is constant; in practice, for the given $z_{IN}$ and $\lambda_{pert}$ in this panel, $\delta \gamma_{IN,\;OD}$ is $\delta$-dominated.
Considering now the middle and right panel in Fig.~(\ref{fig: PertAmp}), $z_{IN}$ and $\lambda_{pert}$ impact the amplitude of $\delta^{(1)}$, while $A_{pert}$, the amplitude of $\mathcal{R}_c$, is constant in these panels. 
Then, when $|\mathcal{R}_{c,\;OD}|>|\delta^{(1)}_{OD}|$, in the $\mathcal{R}_c$-dominated regime (at large $z_{IN}$ and $\lambda_{pert}$) $\delta \gamma_{IN,\;OD}$ shows a plateau, while $\delta \gamma_{IN,\;OD}\propto a(\tau)\lambda_{pert}^{-2}$ in the $\delta$-dominated regime, when $|\mathcal{R}_{c,\;OD}|<|\delta^{(1)}_{OD}|$.

Intuitively, in an OD region ($\delta>0$ and $\mathcal{R}_{c}<0$) you would expect the volume to be smaller than the background average, meaning that $\delta\gamma$ is negative, as that region of space is more compact. 
However in the $\mathcal{R}_c$-dominated regime, $|\mathcal{R}_{c,\;OD}|>|\delta_{OD}|$, the volume element is larger than that of the background in the OD, $\delta\gamma_{OD}>0$. 
This counter-intuitive behaviour is observed when:
\begin{equation}\label{eq: Long_wavelength}
    \lambda_{phy} > \frac{2\pi}{H\sqrt{3F}}.
\end{equation}
This $\mathcal{R}_c$-dominated regime then occurs when the wavelength is much bigger than the Hubble horizon ($>c/H$), so we also call it the long-wavelength regime. 
This phenomenon has previously been discussed \cite{V.F.Mukhanov_etal_1997, L.R.W.Abramo_etal_1997, G.Geshnizjani_R.Brandenberger_2002, R.H.Brandenberger_2002}, where long wavelength modes were proposed to be acting as a form of cosmological constant.
Note that $\delta\gamma$ \textit{per-se} is not a gauge-invariant quantity, rather the $\delta\gamma$ in the synchronous-comoving gauge we are using is the value that the gauge-invariant quantity corresponding to $\delta\gamma$ would have in this gauge.

\section{Code description and Numerical implementation} \label{sec: code description and numerical implementation}
In numerical relativity \cite{M.Alcubierre_2008, T.W.Baumgarte_S.L.Shapiro_2010, M.Shibata_2015}, Einstein's field equations are separated into constraint equations and evolution equations.
So to run simulations an initial spacetime and matter distribution satisfying the constraints is set, then evolved according to the evolution equations, and the constraint equations are used to monitor accuracy throughout the evolution.
While the initial quantities can be set using the ADM formalism \cite{R.Arnowitt_etal_2008, J.W.Jr.York_1979}, in this formalism the evolution equations take a form that is not strongly hyperbolic, this will then cause stability issues in the simulation. 
These quantities need to be transformed to a formulation where the evolution equations are expressed in a strongly hyperbolic form, such as BSSNOK \cite{T.Nakamura_etal_1987, M.Shibata_T.Nakamura_1995, T.W.Baumgarte_S.L.Shapiro_1998}. 
The quantities associated with the fluid that are sourcing Einstein's evolution equations are called the primitive hydrodynamics variables, these are evolved with the conservation equations $\nabla_{\mu} T^{\mu\nu} = 0$. 
Typically these variables are also transformed, in this case to the corresponding conserved quantities see e.g.\cite{E.Bentivegna_2017}, according to the Valencia formulation \cite{J.A.Font_2003, M.Alcubierre_2008}, such that high-resolution shock-capturing numerical schemes can be applied to the evolution equations. 
This is particularly relevant to turbulent scenarios and so are not applied here.

For our simulations we use the open-source code Einstein Toolkit \cite{F.Loffler_etal_2012, S.R.Brandt_etal_2020}. 
This code is a compilation of multiple modules, named thorns, that communicate within the Cactus framework \cite{T.Goodale_etal_2003}. 
These thorns have different tasks and capacities and may be written in C++ or Fortran adapted to Cactus code or in Mathematica or Python to then be converted to C++ Cactus code by Kranc \cite{S.Husa_etal_2006} or NRPy+ \cite{I.Ruchlin_etal_2018}. 
To manage this infrastructure, the simfactory job manager \cite{M.W.Thomas_E.Schnetter_2010} is used for compilation and running jobs.

The initial distributions for our simulations are calculated by our new thorn ICPertFLRW \cite{R.L.Munoz_2023_ICPertFLRW}, developed in Fortran and adapted to Cactus code for this project. 
It defines the initial ADM variables: $\gamma_{ij}$ Eq.~\eqref{eq: metricRc}, $K_{ij}$ Eq.~\eqref{eq: curvRc}, with $\alpha=1$, $\beta^i = 0$ and $\rho$ given by Eq.~\eqref{eq: rhoHam}. 
As explained in Section \ref{sec: fully nonlinear initial conditions}, defining $\rho$ using the Hamiltonian constraint implies that this is initially automatically satisfied, while the momentum constraint is initially satisfied at first-order.
ICPertFLRW then provides the ADM quantities to the ADMBase \cite{F.Loffler_etal_2012} and CT\_Dust thorns \cite{E.Bentivegna_2017}. 
The variables are provided on a Cartesian grid, supported by Carpet \cite{E.Schnetter_etal_2004}; this has mesh refinement capacities although we have not used these in this paper.

To evolve the geometrical variables they are transformed into the BSSNOK formalism \cite{T.Nakamura_etal_1987, M.Shibata_T.Nakamura_1995, T.W.Baumgarte_S.L.Shapiro_1998} and the subsequent variables are evolved by the ML\_BSSN thorn \cite{D.Brown_etal_2009}. 
The primitive hydrodynamics variables are transformed to their conserved form and evolved by CT\_Dust \cite{E.Bentivegna_2017} without hock-capturing schemes. 
They are all integrated with the $4^{th}$ order Runge-Kutta scheme provided by the MoL thorn \cite{F.Loffler_etal_2012}. 
The coupling between the metric and the matter field is ensured by the TmunuBase thorn \cite{F.Loffler_etal_2012}. 

The simulations were run on the Sciama HPC Cluster \cite{Sciama} with box sizes of $32^3$, $64^3$ and $128^3$ data points. 
Sciama's job manager Slurm \cite{Slurm} was made to communicate with simfactory \cite{M.W.Thomas_E.Schnetter_2010}.

\section{Simulation results} \label{sec: Results}
In this section we describe two simulations with the initial conditions of Section \ref{sec: Th framework}, one with $\Lambda$, and one without. 
Both are compared to the spherical collapse model in Section \ref{sec: Peak of the over-density}, and the simulation with $\Lambda$ is then described more in the following subsections.
We fix some of the parameters as in \cite{E.Bentivegna_M.Bruni_2016}, namely $\lambda_{phy,\;IN}=4/H_{IN}=6$Mpc and $\delta_{IN,\; OD} = 3 \times 10^{-2}$, where we assume $H_0 = c h / 2997.9$ Mpc${}^{-1}$, with $c=1$ and $h=0.6737$ \cite{Planck_CMB_2018}.
As such the simulation without $\Lambda$ starts at $z_{IN}=205.4$ with $\lambda_{pert}=1206$Mpc and the simulation with $\Lambda$ at $z_{IN}=302.5$ with $\lambda_{pert}=1821$Mpc. 
The initial $\delta_{OD}$ is chosen in order for the OD to collapse at $2 < z < 5$.
These initial conditions are evolved up until the OD collapses on itself, in practice the simulation `crashes' as NaN\footnote{ Not a Number.} values appear. 
This is due to our fluid description of matter and use of synchronous-comoving coordinates, while such a structure would otherwise be expected to relax into a virialised dark matter halo.

\subsection{Over-density peak evolution and Top-Hat model} \label{sec: Peak of the over-density}
\begin{figure}[th]
    \centering
    \includegraphics[width=\linewidth]{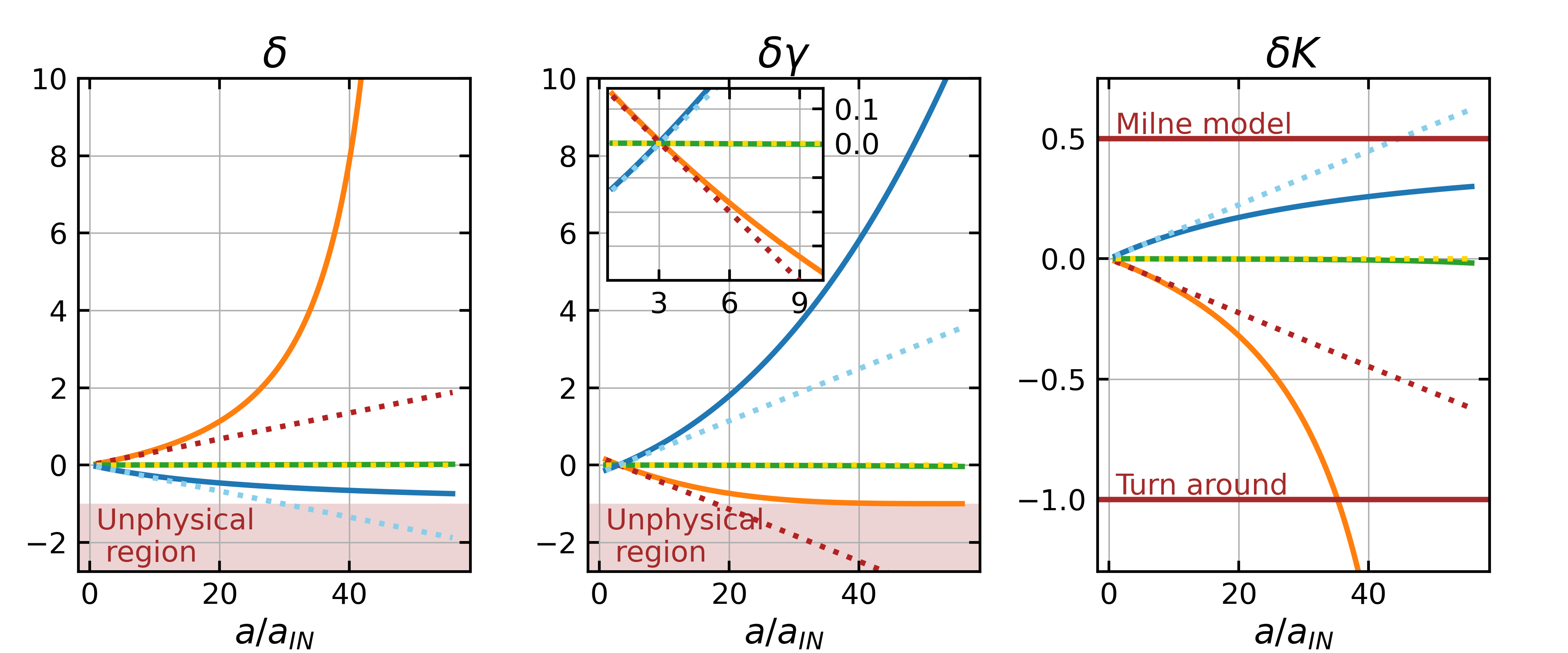}
    \includegraphics[width=\linewidth]{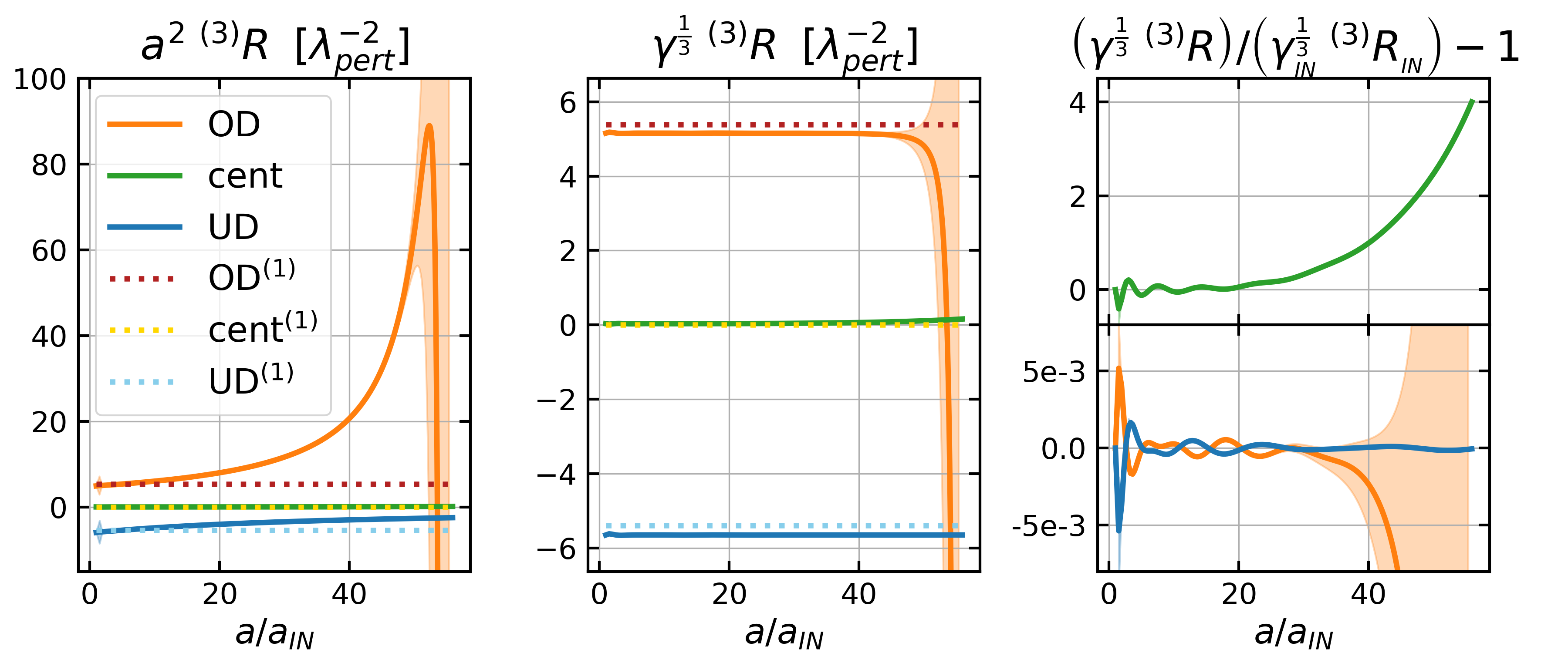}
    \caption{Evolution of various quantities at the peak of the over and under-density (OD in orange and UD in blue) as well as the central location of the simulation box (in green). 
    \textit{Top:} the matter density, volume, and expansion contrasts $\delta$, $\delta\gamma$ and $\delta K$. 
    \textit{Bottom:} the conformal 3-Ricci scalar defined with the $\Lambda$CDM FLRW scale factor $a^2{}^{(3)}R$; conformal 3-Ricci scalar defined with the nonlinear scale factor $\gamma^{\frac{1}{3}}{}^{(3)}R$; the same quantity normalised with its initial value $\left( \gamma^{1/3} {}^{(3)}R \right) / \left( \gamma^{1/3}_{{}_{IN}} {}^{(3)}R_{{}_{IN}} \right) - 1$.
    The dashed lines are the first-order projections from Eq.~(\ref{eq: delta1}), Eq.~(\ref{eq: delta_gamma1_K1}) and  Eq.~(\ref{eq: Rc_comoving_curvature_perturbation}), and the full lines are the simulation results. Initial conditions are $\delta_{IN, \; OD} = 3 \times 10^{-2}$, $z_{IN} = 302.5$ and $\lambda_{pert} = 1821$Mpc, and $\Lambda$ is present. 
    Error bars, when visible, are indicated as shaded regions.}
    \label{fig: Collapse_evo}
\end{figure}

The evolution of the inhomogeneities at the peak of the OD and at the bottom of the UD is presented in Fig.~(\ref{fig: Collapse_evo}) for the $\Lambda$CDM case. 
For the top row, from left to right, we show: the density contrast $\delta$, the volume contrast $\delta\gamma$ and the expansion contrast $\delta K$ in Eq.~(\ref{eq: pert}). 
The dashed lines are the first-order expectations from Eq.~(\ref{eq: delta1}), Eq.~(\ref{eq: delta_gamma1_K1}) and  Eq.~(\ref{eq: Rc_comoving_curvature_perturbation}) while the full lines are the results of the simulation. 
The separation between those lines shows a departure from linearity, which happens early on in the simulation. 
The unphysical regions ($\rho$ and $\gamma$ need to be positive) and Milne model limit ($a\propto t$ \cite{N.Vittorio_2018}) in the plots show that these departures from linearity are indeed in a sense necessary for this system to remain physical. 

In the centre of the OD, still on the top row from left to right: $\delta_{OD}$ becomes very large, the volume element tends towards zero, so that $\delta\gamma_{OD} \rightarrow -1$, the initial expansion is more and more decelerated until it turns around (TA) and contraction begins, when $K_{OD} = 0$ and $\delta K_{OD} = -1$.
The reverse is observed in the centre of the UD: the density tends to zero $\delta_{UD} \rightarrow -1$, the volume element becomes much larger than the reference FLRW and the expansion is faster.
In the centre of the simulation box, where initially $\mathcal{R}_c=0$, the first-order quantities all remain zero, but the non-linearity introduced by ${}^{(3)}R$ in the initial conditions makes all quantities in the figure measurably non-zero (beyond numerical error) although they remain very small.

Notice the sign change in the volume contrast $\delta\gamma$ at $a/a_{IN} \simeq 3.1$. 
This behaviour is representative of the transition experienced by long wavelength perturbations as they evolve from the $\mathcal{R}_c$-dominated to $\delta$-dominated regime, according to Eq.~(\ref{eq: Long_wavelength}) \cite{V.F.Mukhanov_etal_1997, L.R.W.Abramo_etal_1997, G.Geshnizjani_R.Brandenberger_2002, R.H.Brandenberger_2002}. 

\begin{table}[ht]
\centering
\begin{tabular}{cc||c|c|c|c}
 & & \multirow{2}{*}{\centering Top-Hat, $\Lambda=0$} & \multirow{2}{*}{\centering Here, $\Lambda=0$} & \multirow{2}{*}{\centering Here, $\Lambda\neq0$} & \multirow{2}{2cm}{\centering E.B. \& M.B. (2016) \cite{E.Bentivegna_M.Bruni_2016}} \\
 &&&&&\\
\hline
\hline
Initially & $z_{IN}$ & & 205.4 & 302.5 & 205.4 \\
\hline
            & $a/a_{IN}$ & 35.4137 & 35.24467 $\pm$ 7e-5 & 35.195 $\pm$ 3e-3 & 60 \\
            & $z$ & & 4.85620 $\pm$ 1e-5 & 7.6234 $\pm$ 7e-4 & 2.44 \\
Turn Around (TA)& $\gamma^{1/6}_{OD}/\gamma^{1/6}_{IN,\;OD}$ & & 20.10169 $\pm$ 3e-5 & 20.0600 $\pm$ 1e-4 & \\
$K=0$       & $\langle \gamma^{1/6}\rangle_{\mathcal{D}}/\langle \gamma^{1/6}\rangle_{\mathcal{D},\;IN}$ & & 35.2064 $\pm$ 1e-4 & 35.154 $\pm$ 3e-3 & \\
            & $\delta^{(1)}_{OD}$ & 1.06241 & 1.05734 $\pm$ 2e-6 & 1.05584 $\pm$ 8e-5 & 1.8* \\
            & $\delta_{OD}$ & 4.55165 & 4.55164 $\pm$ 1e-5 & 4.5626 $\pm$ 5e-4 & \\
\hline
              & $a/a_{IN}$ & 56.22 & 55.9 $\pm$ 1e-1 & 55.87 $\pm$ 8e-2 & 96 \\
              & $z$ & & 2.692 $\pm$ 7e-3 & 4.432 $\pm$ 8e-3 & 1.15 \\
Collapse & $\gamma^{1/6}_{OD}/\gamma^{1/6}_{IN,\;OD}$ & & 0.4 $\pm$ 6e-1 & 0.8 $\pm$ 2e-1 & \\
      /Crash & $\langle \gamma^{1/6}\rangle_{\mathcal{D}}/\langle \gamma^{1/6}\rangle_{\mathcal{D},\;IN}$ & & 55.8 $\pm$ 1e-1 & 55.77 $\pm$ 2e-2 & \\
              & $\delta^{(1)}_{OD}$ & 1.686 & 1.678 $\pm$ 3e-3 & 1.676 $\pm$ 2e-3 & 2.88 \\
              & $\delta_{OD}$ & $+\infty$ & 2e+6 $\pm$ 2e+6 & 4e+5 $\pm$ 4e+5 & \\
              \hline
              Virialisation & $a/a_{IN}$ & 52.64 & 52.5055 $\pm$ 9e-4 & 52.469 $\pm$ 2e-3 &  \\
              $R=R_{TA}/2$ & $\delta_{OD}$ & 145.84 & 145.84 & 145.84 & \\
              \hline
              Virialisation & $a/a_{IN}$ & 56.22 & 52.83625 $\pm$ 7e-5 & 52.801 $\pm$ 2e-3 &  \\
            $R=R_{TA}/2$ \& $\tau=\tau_{C}$ & $\delta_{OD}$ & 176.65 & 176.65 & 176.65 & \\
\end{tabular}
\caption{Various variables during the evolution of an over-density (OD) whose initial (IN) density contrast is $\delta_{IN,\; OD} = 0.03$ and physical size $\lambda_{phy,\; IN} = 4 / H_{IN}$. 
These variables are recorded for four scenarios at different \textit{stages of the evolution}: the turn around (TA), the collapse/crash of the OD, and its virialisation according to two different definitions, when the radius of the Top-Hat sphere is half its radius at TA, and when that property happens at the time of the collapse. 
\textit{The four scenarios} are the theoretical Top-Hat spherical and homogeneous collapse model (first column \cite{J.E.Gunn_J.R.Gott_1972, J.A.Peacock_1999, H.Mo_etal_2010, N.Vittorio_2018}) and three numerical relativity simulations of a 3-D sinusoidal peak. 
These are: our simulations with a purely growing mode with $\Lambda = 0$ (second column), and with $\Lambda \neq 0$ (third column); from \cite{E.Bentivegna_M.Bruni_2016}, with a growing and decaying mode with $\Lambda = 0$ (fourth column). 
\textit{The variables} are: the normalised background scale factor $a / a_{IN}$, with its corresponding redshift $z$ and linear density contrast $\delta^{(1)}_{OD}$ ($\delta^{(1)}_{OD} = \delta_{IN, \; OD} \; a / a_{IN}$ for EdS), this is to be compared to the local scale factor $\gamma^{1/6}_{OD} / \gamma^{1/6}_{IN, \; OD}$, the domain average scale factor $\langle \gamma^{1/6} \rangle_{\mathcal{D}} / \langle \gamma^{1/6} \rangle_{\mathcal{D},\; IN}$ (averaged over the whole simulation box), and the nonlinear density contrast $\delta_{OD}$. 
For the two definitions of virialisation $a/a_{IN}$ is recorded at the given $\delta_{OD}$. 
The asterisk indicates a factor of three correction to the value reported in \cite{E.Bentivegna_M.Bruni_2016}.}
\label{tab: Collapse}
\end{table}

Then the second row of panels in Fig.~(\ref{fig: Collapse_evo}) show, first on the left, the conformal 3-Ricci scalar defined with respect to the $\Lambda$CDM FLRW scale factor, $a^2{}^{(3)}R$ \cite{M.Bruni_etal_2014_Mar, M.Bruni_etal_2014_Sep}. 
At first order this quantity is conserved at all scales for dust, as shown by the dashed lines, however in the OD the curvature is positive and grows larger and larger up until the crash, while in the UD it is initially negative and tends towards zero. 
The middle panel on the other hand shows the conformal 3-Ricci scalar defined with respect to the nonlinear scale factor from the simulation, $\gamma^{\frac{1}{3}} {}^{(3)}R$: for the OD, essentially this is conserved throughout the evolution up until just before the crash.
Indeed when normalised with its initial value, as can be seen in the rightmost panel, only sub-percent fluctuations are observed in the UD and OD (when error bars are reasonable), but a more notable deviation can be seen in the central location. 
This shows that the locations at the top/bottom of the inhomogeneity conserve their local nonlinear conformal curvature, which is essentially consistent with the closed FLRW description of the Top-Hat model.
As the volume element in the OD shrinks, the curvature grows, therefore the two effects evolve together such that nonlinear conformal curvature is constant, conversely in the UD the volume element grows and the curvature tends towards zero such that the conformal curvature is also constant. 
In the central region the volume element shrinks and the curvature grows like in the centre of the OD, although these deviations are too small to be seen in Fig.~(\ref{fig: Collapse_evo}); however in this location the nonlinear conformal curvature is not conserved. 
This may be due to this location having a much greater density gradient $\partial_i \mathcal{R}_c = A_{pert} k_{pert}$ than the OD and UD centre $\partial_i \mathcal{R}_c = 0$.

The exact values of various quantities at TA, at times corresponding to virialisation according to two different definitions \cite{J.E.Gunn_J.R.Gott_1972, J.A.Peacock_1999, H.Mo_etal_2010, N.Vittorio_2018}, and at the collapse/crash time are listed in Table~(\ref{tab: Collapse}). 
Defining $R$ as the radius of the Top-Hat sphere, in this model $R$ increases to reach its maximal size at TA, $R_{TA}$, when $K$ changes sign, from expansion to contraction, so TA measurements are taken when $K = 0$.  
At TA the kinetic energy is zero, $E_{Kin,\;TA} = 0$ and so the total energy is contained in the potential energy $E_{Tot} = E_{Pot,\;TA} \propto 1 / R_{TA}$.
After that, $R$ shrinks and collapses to $R = 0$.
While the Top-Hat model does not have the mechanisms to enable virialisation, there are two different definitions typically used to approximate it. 
Virialisation happens when the potential energy is double the kinetic energy, with a sign change, $E_{Pot,\;V} = - 2 E_{Kin,\;V}$.
As energy is conserved, this means that the potential energy at virialisation can be related to the potential energy at TA, $E_{Pot,\;V} = 2 E_{Pot,\;TA}$, therefore at virialisation the radius becomes $R_{V} = R_{TA} / 2$.
The first definition of virialisation is then when $R$, evolving according to the Top-Hat model, reaches $R_{TA} / 2$ \cite{J.A.Peacock_1999}.
The second definition also works with $R_{V} = R_{TA} / 2$ but assumes that relaxation mechanisms are present, and so establishes that $R$ would reach this value at the time of the collapse $\tau_{C}$ \cite{J.A.Peacock_1999, N.Vittorio_2018}.
This means that this second definition has a discontinuity in the $R$ evolution, which is assumed to be filled with relaxation mechanisms.
Either way, these two definitions predict specific nonlinear $\delta_{OD}$, so here we record $a/a_{IN}$ when $\delta_{OD}$ reaches those values.
Some of the values reported in Table~(\ref{tab: Collapse}) are related to times between recorded iterations, so they were obtained with a linear fit.
Then, for the collapse/crash, the last valid values of $a/a_{IN}$ and $\delta_{OD}$ are recorded.

In our simulations, the TA and collapse/crash, with and without $\Lambda$, occur at an earlier time than the time in \cite{E.Bentivegna_M.Bruni_2016}. 
This shows that the presence of the decaying mode in their case has significantly slowed down the evolution, as was also shown by \cite{W.E.East_etal_2018}. 
Correspondingly, they also have a bigger\footnote{That is, for the linearly extrapolated density contrast we have $\delta^{(1)}_{TA, \; OD} = 1.8$ for a TA at $a/a_{IN} = 60$ as in \cite{E.Bentivegna_M.Bruni_2016}, thus correcting the value for $\delta^{(1)}$ at TA reported in \cite{E.Bentivegna_M.Bruni_2016}, $\delta^{(1)}_{T} = 0.6$. Similarly, given that the collapse in \cite{E.Bentivegna_M.Bruni_2016} is at $a/a_{IN} \simeq 96$,  $\delta^{(1)} \simeq 0.96$ under the same assumptions, while the correct value is $\delta^{(1)} \simeq 2.8$, as we report in Table~(\ref{tab: Collapse}). The presence of the decaying mode in \cite{E.Bentivegna_M.Bruni_2016}  implied that a direct match with the prediction of the Top Hat model was not expected and somehow confused the interpretation of the results. This was based on assuming that the initial density contrast was $\delta_{IN, \; OD}=\delta_i=10^{-2}$, as reported in the text around Eq.\ (9) in \cite{E.Bentivegna_M.Bruni_2016}, while the correct value of the initial $\delta$ was $\delta_{IN, \; OD} = 3\delta_i = 3\times 10^{-2}$, as it is clearly visible in the leftmost panel of Fig.\ 1 and their Eq.\ (9).} $\delta^{(1)}_{OD}$ at those moments, this is simply due to the longer evolution since $\delta^{(1)}_{OD} = \delta^{(1)}_{IN, \; OD} \; a/a_{IN}$ in EdS. 

Otherwise, we see that at the peak of the OD we reach TA and collapse/crash precisely when the Top-Hat model predicts it, with the expected $a/a_{IN}$, $\delta^{(1)}_{OD}$ and $\delta_{OD}$ values in agreement with \cite{W.E.East_etal_2018}. 
With the conservation of the local conformal curvature, this shows that the Top-Hat model provides excellent predictions for the centre of the OD.
Furthermore, the domain averaged scale factor, $\langle \gamma^{1/6}\rangle_{\mathcal{D}}/\langle 
\gamma^{1/6}\rangle_{\mathcal{D},\;IN}$, is also close to the Top-Hat model prediction for $a/a_{IN}$. This is not the case for the local measurement, $\gamma^{1/6}_{OD}/\gamma^{1/6}_{IN, \; OD}$, which instead shows the compactness of the region.

For virialisation, we recover the expected $a/a_{IN}$ for the first definition of $R = R_{TA} / 2$, but not for the second $R = R_{TA} / 2$ and $\tau = \tau_{C}$. 
The first definition is based on the exact evolution of $R$ for the Top-Hat model, while the second provides an approximation by making the assumption that, relaxation mechanisms are present. 
The matter in these simulations is described as a pressureless perfect fluid, it therefore does not have any relaxation mechanism, so instead, as we observe in the centre of the OD, the evolution of the density contrast is well predicted by the Top-Hat model.

We see a slight difference depending on the presence of $\Lambda$ in the simulation. However, the error estimates overlap in many cases and we measure up to a maximum $\simeq 0.57 \%$ difference between the $\Lambda = 0$ and the $\Lambda\neq 0$ simulations.

\subsection{Raychaudhuri equation: local evolution and Top-Hat approximation} \label{sec: Raychaudhuri equation}
\begin{figure}[t]
    \centering
    \includegraphics[width=0.5\linewidth]{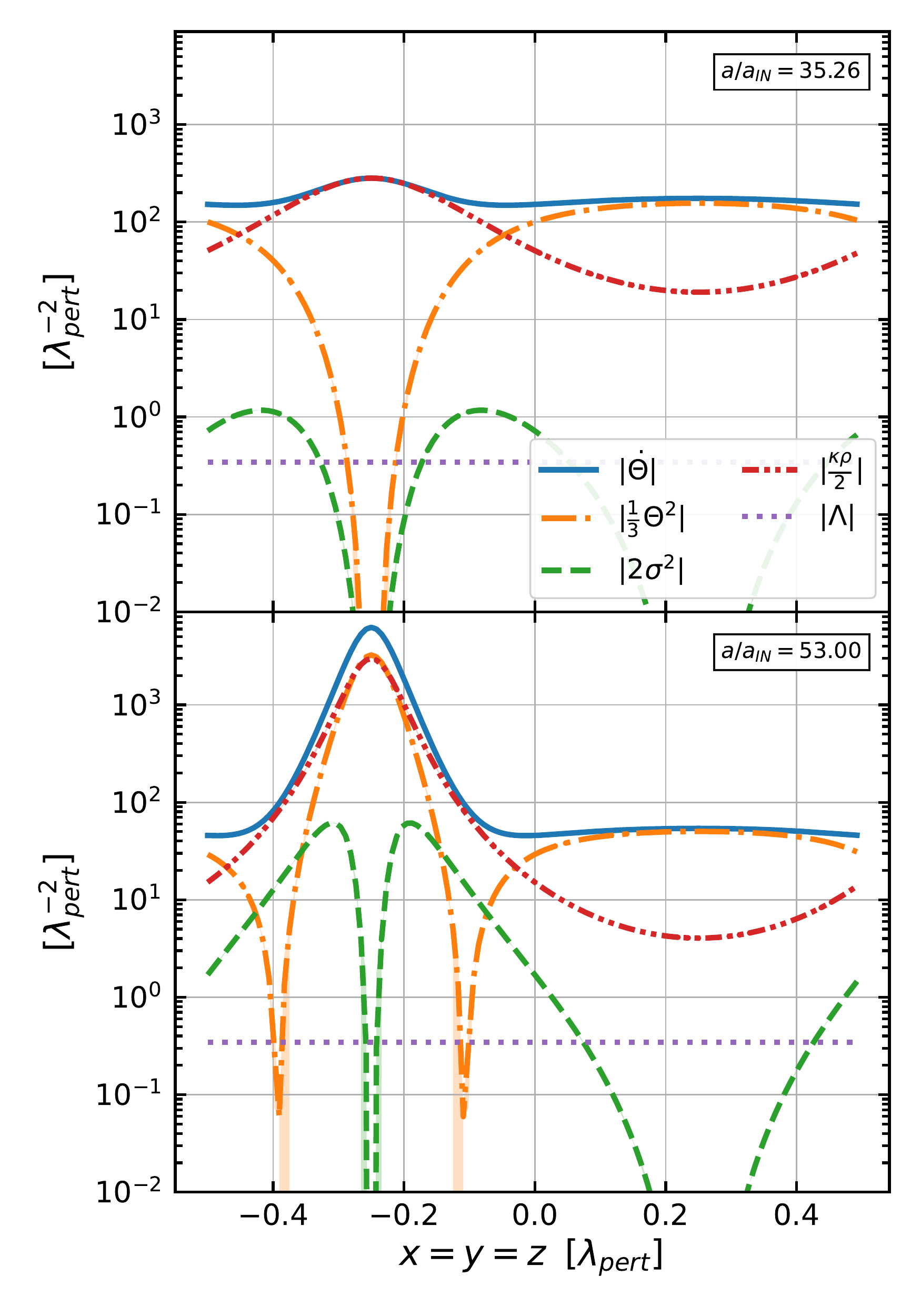}
    \caption{Contributions to the Raychaudhuri equation just after the turn-around of the peak (top panel) and just before the crash (bottom panel): since $c=G=1$ all these terms have units of length$^{-2}$, therefore we measure them in $\lambda_{pert}^{-2}$ units. Each term is presented along the $x=y=z$ diagonal of the data box, the peak of the over-density is at $x=-0.25\lambda_{pert}$ and the bottom of the under-density is at $x=0.25\lambda_{pert}$. Error bars, when visible, are indicated as shaded regions.}
    \label{fig: Ray_slice}
\end{figure}
Our results, in either case, show that at the peak of the OD the Top-Hat model is an excellent approximation. 
To understand this, consider the Raychaudhuri equation (\ref{eq: Raychaudhuri}) describing the local evolution of the fluid expansion scalar.
Each term contributing to $\dot{\Theta}$ is plotted along the $x=y=z$ diagonal, in Fig.~(\ref{fig: Ray_slice}). 
This direction goes from the centre of the OD through the centre of the face of the octahedron such that it also goes through the centre of the UD (this is the dashed green line in Fig.~(\ref{fig: IniDelta1d})). 

The matter density $\rho$ curve, i.e.\ the dot-dot-dashed red line in Fig.~(\ref{fig: Ray_slice}), clearly shows the OD and UD regions located at $\pm 0.25 \lambda_{pert}$. 
The shear contribution, $\sigma^2$, shown with the dashed green line, is subdominant everywhere; it does grow around the OD but it is always essentially zero at the peak of the OD and at the centre of the UD. 
The reason that $\sigma^2$ is negligible in these specific locations is because of the triaxial symmetry, so that around these two points the distribution is almost spherical.
The fact that the shear gives a negligible contribution to the Raychaudhuri equation implies that at the OD and the UD locations the evolution is in essence independent of the environment.   
Mathematically, neglecting the shear implies that the Raychaudhuri equation is only coupled to the continuity equation \eqref{eq: continuity}: then at the OD these two equations are formally identical to those in FLRW with positive 3-curvature, as implied by the Hamiltonian constraint \eqref{eq: Hamiltonian}.
Therefore, at the peak, the Top-Hat model is a very good approximation.

Then the expansion, $\Theta$, shown with the dot-dashed orange line, peaks downwards, $\Theta=-K=0$, in locations experiencing TA. 
The peak of the OD experiences TA first, then its surrounding region. 
This identifies the infalling domain discussed in the next Section~\ref{sec: infalling domain}. 

\subsection{Expansion of the infalling domain} \label{sec: infalling domain}
\begin{figure}[t]
    \centering
    \includegraphics[width=\linewidth]{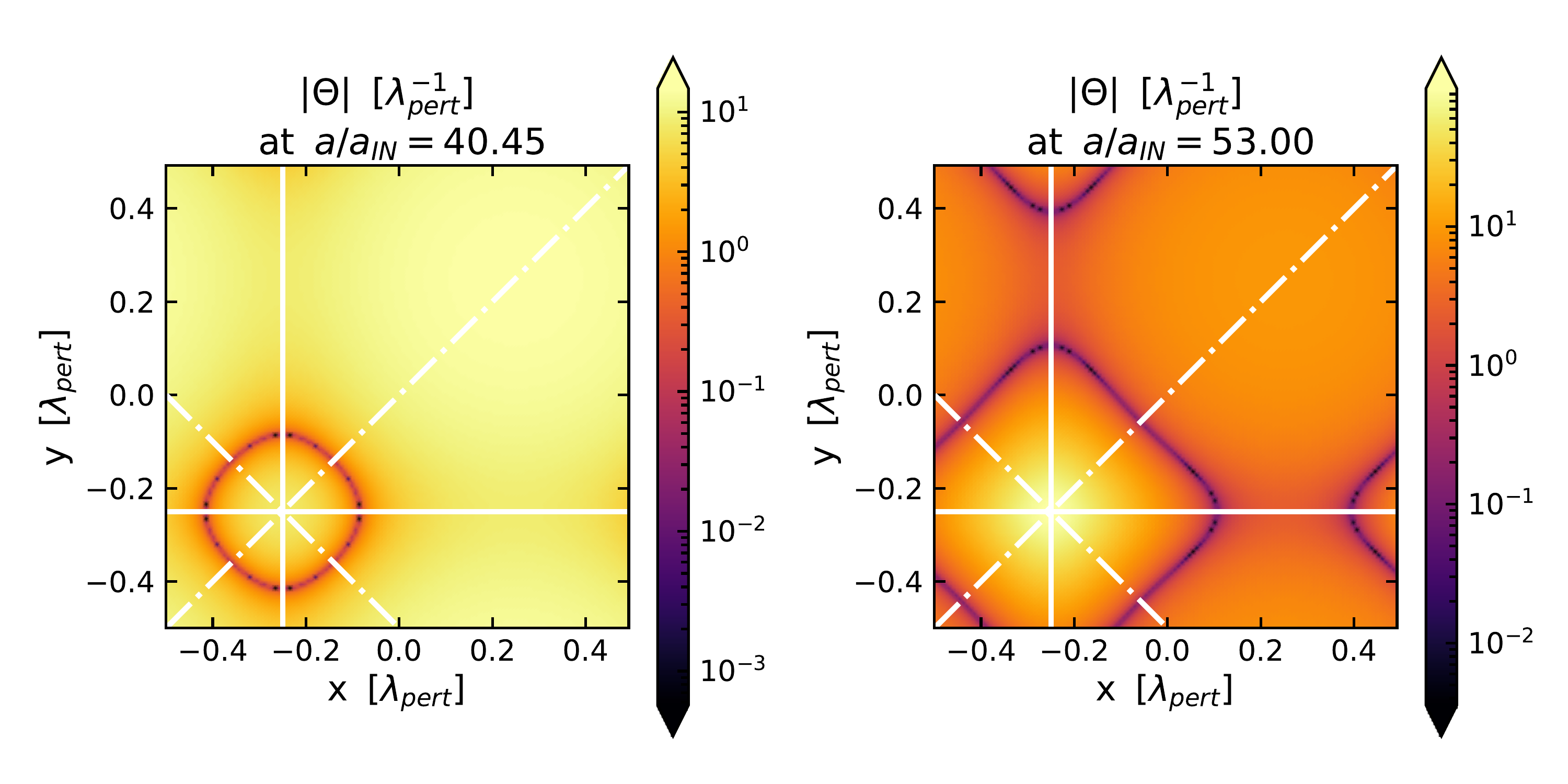}
    \caption{Absolute expansion scalar $\Theta$ in units $\lambda_{pert}^{-1}$ in the x-y plane passing by the peak of the over-density ($z=-0.25\lambda_{pert}$) at $a/a_{IN}=40.45$ and $53.00$. The full lines indicate directions along the vertices and the dash-dotted lines are the directions along the centre of the edges.}
    \label{fig: TAradius_evo_2d}
\end{figure}
Throughout the evolution of the collapsing region, the expansion $\Theta=-K$ of the OD is positive but more decelerated than the reference $\Lambda$CDM, until it reaches TA at $\Theta=0$ and then contracts inwards $\Theta<0$. 
The peak of the OD is the first to reach TA, followed by its surrounding region, where points at a larger distance from the peak reach TA at later times. 

The infalling region, identified using the TA boundary $\Theta=0$, is shown in Fig.~(\ref{fig: TAradius_evo_2d}) at two different times. 
Initially, the boundary surface is close to spherical symmetry, but later, as it encompasses a greater comoving volume and therefore a larger mass, the non-spherical shape becomes apparent. 
As the TA boundary expands outward it tends towards an octahedron, this appears as an almost square boundary in the 2-D slicing through the box in the right panel of Fig.~(\ref{fig: TAradius_evo_2d}), extending beyond the box sides with the periodic boundary condition. 

\begin{figure}[th]
    \centering
        \includegraphics[width=\linewidth]{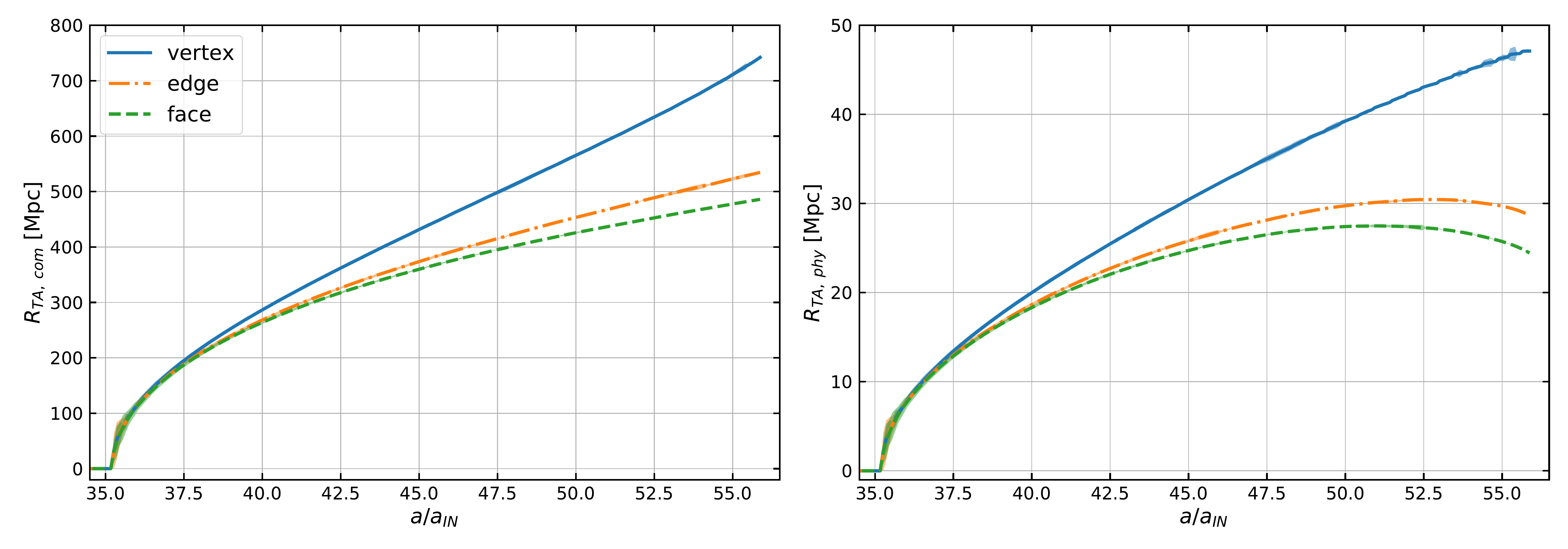}
    \caption{Evolution of the turn around radius $R_{TA}$ - distance from the peak of the over-density to $\Theta=0$ in three directions. On the left, $R_{TA}$ measured in terms of the comoving length today; on the right the corresponding proper length; we emphasise that the physical length is an order of magnitude smaller than the comoving length. Error bars, when visible, are indicated as shaded regions.}
    \label{fig: TAradius_evo_1d}
\end{figure}

With octahedrons, there are three directions of interest: from the centre to the vertices, to the centre of the edges, and to the centre of the faces. 
The plane in Fig.~(\ref{fig: TAradius_evo_2d}) shows the vertex and the centre of the edge directions (full and dash-dotted lines).
As the TA boundary $\Theta=0$ expands outward, we measure the distance between the peak of the OD and the TA point in each direction, which we call the TA radius $R_{TA}$. 
The evolution along the three different directions is presented in Fig.~(\ref{fig: TAradius_evo_1d}), where we depict the comoving coordinate TA radius $R_{TA,\;com}$ in the left panel, and the physical TA radius $R_{TA,\;phy}$ in the right panel, see Appendix~\ref{sec: Num_int}. 

In the left panel, the TA boundaries grow in the same way in the three directions, so long as they stay in the region that is almost spherically symmetric around the peak, and then they split out according to the direction-dependent distribution. 
In the directions with the biggest $\delta$, the TA radius grows the fastest.

This is also true when we consider the proper distances $R_{TA,\;phy}$, by integrating with the local scale factor, see Appendix~\ref{sec: Num_int}, which are shown in the right panel of Fig.~(\ref{fig: TAradius_evo_1d}). 
Notably, we see that in the two directions that go through an UD region, edges and faces, $R_{TA,\;phy}$ stops growing and starts decreasing. 
So in these two directions, the region of infalling material reaches a maximal size and then starts shrinking, while in the direction where $\delta$ is always positive, the infalling region continues to grow. 

\subsection{Evolution of a comoving sphere} \label{sec: Comoving sphere}
\begin{figure}[th]
    \centering
        \includegraphics[width=\linewidth]{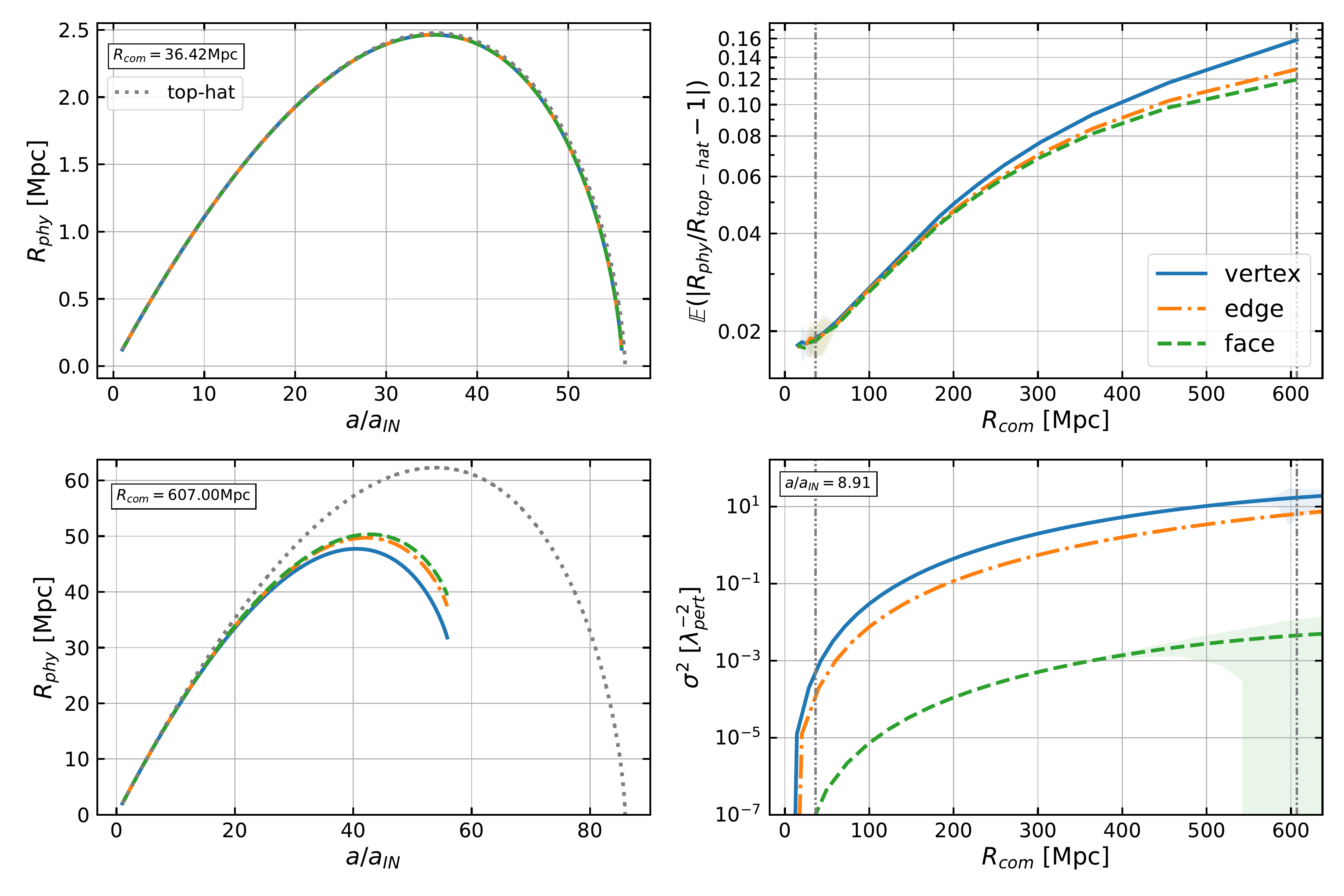}
    \caption{\textit{Left panels:} evolution of proper physical radius of two comoving spheres, one small (top panel) and one large (bottom panel), centred on the peak of the over-density, in all three directions, compared to the Top-Hat spherical and homogeneous collapse model.
    The comoving radii are listed as text in the plots. 
    The Top-Hat models were computed using the domain average $\delta$ within the two spheres, $\langle \delta \rangle_{\mathcal{D}}$ see Appendix~\ref{sec: Num_int}. 
    \textit{Top right panel:} average relative difference between the simulation results and the Top-Hat model prediction for a range of comoving radii. 
    The two cases on the left are identified with grey dot-dot-dashed lines.
    \textit{Bottom right panel:} shear in the three directions from the peak of the over-density.
    Error bars, when visible, are indicated as shaded regions.}
    \label{fig: Comoving_sphere}
\end{figure}

We can draw another comparison to the Top-Hat model by considering the evolution of a comoving sphere, a region with constant mass, centred on the peak of the OD and compare its evolution with that of a homogeneous spherical Top-Hat with $\delta=\langle\delta\rangle_{\mathcal{D}}$. 
For a given comoving radius, we integrate to measure the proper physical radius and present it in the left panels of Fig.~(\ref{fig: Comoving_sphere}). 
Two comoving radii are considered, one small $0.02\lambda_{pert}$, where we see that all three directions behave in the same way, and one big $0.33\lambda_{pert}$, with a direction-dependent evolution such that the bigger the $\delta$, the sooner the collapse. 
In the latter case, we see how a spherical comoving region gradually gets distorted in physical space.

The Top-Hat models, grey dotted lines in the left panels of Fig.~(\ref{fig: Comoving_sphere}), were computed with the domain average $\delta$ within the given comoving sphere, $\langle\delta\rangle_{\mathcal{D}}$, see \cite{R.L.Munoz_2022_sphereint} and Appendix~\ref{sec: Num_int}. 
The small comoving radius case closely follows the Top-Hat model but falls just short of reaching collapse as the peak had already reached that point. 
In the large comoving radius case, there is a clear departure from the Top-Hat model, the region would collapse sooner than what the Top-Hat model would have predicted. 
Indeed for such an inhomogeneity, it is unfair to compare it to a homogeneous sphere.

The average relative difference between the physical radius and the Top-Hat model prediction is measured for a range of comoving radii and presented in the top right panel of Fig.~(\ref{fig: Comoving_sphere}). 
The grey dot-dot dashed vertical lines identify the two cases on the left panels.
This indeed shows that as the radius of the comoving sphere is increased, the bigger the difference between the results and the corresponding Top-Hat model.
This indicates the limit with which inhomogeneous structures can be predicted with homogeneous models.

In Section~\ref{sec: Raychaudhuri equation} we identified the sub-dominance of shear in the proximity of the peak to be the main reason why the evolution of this region closely follows the Top-Hat model prediction described in Section~\ref{sec: Peak of the over-density}.
Then, in the bottom right panel of Fig.~(\ref{fig: Comoving_sphere}) we also show the shear as a function of the comoving radius $R_{com}$.
Indeed, further out from the peak of the OD, the shear is no longer negligible, even if it is still subdominant at this radius as a contribution to the Raychaudhuri equation in Fig.~(\ref{fig: Ray_slice}).

\subsection{Gravito-electromagnetism} \label{sec: gravito-electromagnetism}
\begin{figure}[th]
    \begin{minipage}[b]{0.49\linewidth}
        \centering
        \includegraphics[width=\linewidth]{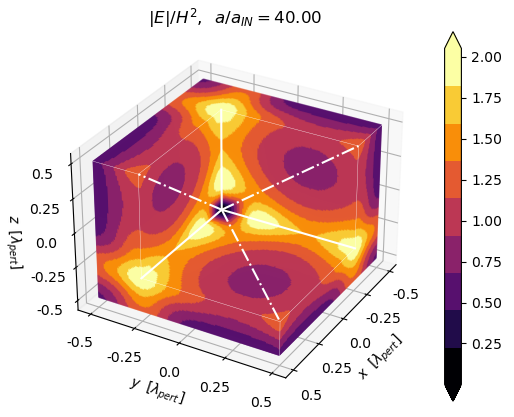}
    \end{minipage}
    \begin{minipage}[b]{0.49\linewidth}
        \centering
        \includegraphics[width=\linewidth]{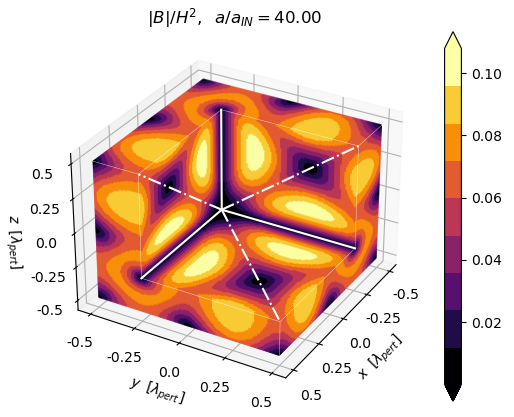}
    \end{minipage}
    \caption{Distribution of the electric and magnetic parts of the Weyl tensor (left and right) in the simulation box, made dimensionless with the Hubble scalar $H$. The $x$, $y$, and $z >-0.25\lambda_{pert}$ region is removed exposing the centre of the over-density. The full white lines go through the vertices and dash-dotted lines through the centre of the edges of an octahedron centred at the over-density.}
    \label{fig: 3d_EandB}
\end{figure}
\begin{figure}[th!]
    \centering
        \includegraphics[width=\linewidth]{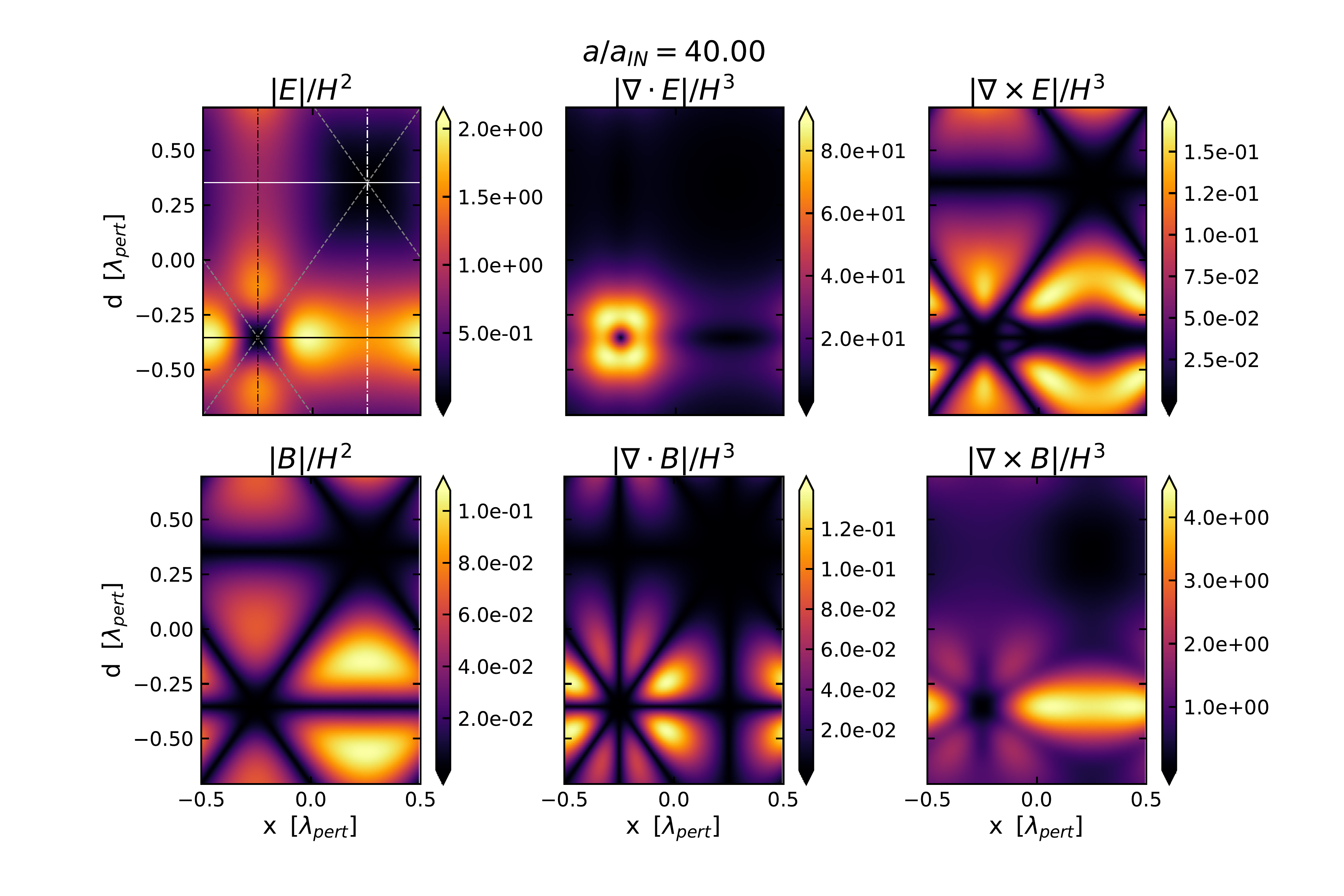}
    \caption{Magnitude of the electric and magnetic parts of the Weyl tensor, and their divergences and curls along the $x$ and $y=z$ plane of the simulation box (with $d^2=y^2+z^2$) at $a/a_{IN}=40.0$, and made dimensionless with the Hubble scalar $H$. 
    The relevant axes of symmetry are marked on the top left panel. 
    The directions going from the centre of the octahedrons to the vertices are marked with full lines, to the centre edges with dash-dotted, and to the centre of the faces with dashed lines. 
    Directions going from the centre of the over-density are marked with black lines, and from the under-density with white lines, the directions going along the faces are valid for both the over-density and the under-density and so are in grey.}
    \label{fig: 2dEB_div_curl}
\end{figure}

The electric and magnetic parts of the Weyl tensor defined with respect to the fluid flow, $u^{\alpha}$, are given by \cite{A.Matte_1953, S.W.Hawking_1966, R.Maartens_B.A.Bassett_1998, G.F.R.Ellis_H.van_Elst_1999, G.F.R.Ellis_2009, G.F.R.Ellis_etal_2012}:
\begin{equation}\label{eq: EBGeometrical}
    E_{\alpha\mu} = u^{\beta} u^{\nu} C_{\alpha\beta\mu\nu},
    \;\;\;\;\;\;\;\;\;
    B_{\alpha\lambda} = u^{\beta} u^{\sigma} \frac{1}{2} C_{\alpha\beta\mu\nu} {\epsilon^{\mu\nu}}_{\lambda\sigma},
\end{equation}
with $\epsilon_{\alpha\beta\mu\nu}$ the Levi-Civita completely antisymmetric tensor fixed with $\epsilon_{0123} = \sqrt{|det(g_{\alpha\beta})|}$. 
$E_{\alpha\beta}$ and $B_{\alpha\beta}$ describe the non-local gravitational field. 
In general, in 3+1 they are computed with respect to a unit time-like hypersurface-orthogonal vector field $n^\alpha$.

\begin{figure}[th]
    \centering
        \includegraphics[width=\linewidth]{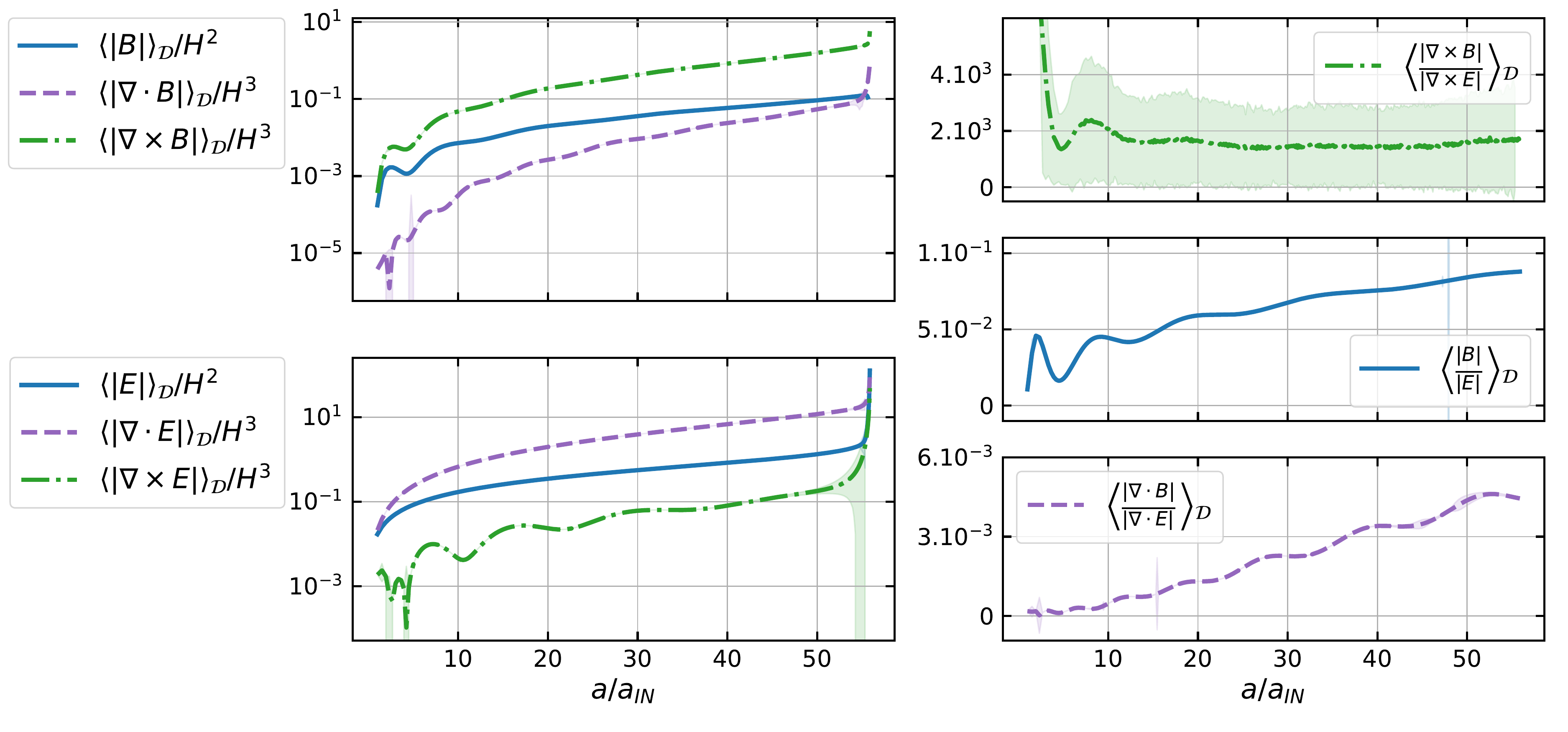}
    \caption{\textit{Top left:} domain average magnitude of the magnetic part of the Weyl tensor, of its divergence and curl throughout the simulation, made dimensionless with the Hubble scalar $H$. \textit{Bottom left:} same as above but with the electric part of the Weyl tensor. \textit{Right:} ratios between the magnetic and electric terms. Error bars, when visible, are indicated as shaded regions.}
    \label{fig: 1dEB_div_curl}
\end{figure}

We compute $E_{\alpha\beta}$ and $B_{\alpha\beta}$ with EBWeyl, the code presented and tested in Paper 1 \cite{R.L.Munoz_M.Bruni_2022, R.L.Munoz_2022_EBWeyl}, together with their divergence $(\nabla\cdot E)_{\alpha}$ and curl $(\nabla\times E)_{\alpha\beta}$ \cite{G.F.R.Ellis_etal_2012} defined in the hypersurface with metric $\gamma_{ij}$, where $n_\mu=\{-\alpha,\;0,\;0,\;0\}$ and $\alpha$ is the lapse function of the 3+1 formalism, see Paper 1 \cite{R.L.Munoz_M.Bruni_2022}:
\begin{equation}\label{eq: div and curl}
    (\nabla\cdot E)_{\mu} = D^{i} E_{i\mu},
    \;\;\;\;\;\;\;\;\;\;
    (\nabla\times E)_{\mu\nu} = -{\epsilon^{\alpha \beta \sigma}}_{(\mu} n_\alpha D_{\beta} E_{\nu)\sigma} = \alpha {\epsilon^{0ij}}_{(\mu} D_{i} E_{\nu)j},
\end{equation}
where $D_{i}$ is the covariant derivative with respect to $\gamma_{ij}$. 
In this paper we use the synchronous-comoving gauge, then the lapse $\alpha=1$, the shift $\beta^i=0$, and the normal to the $\gamma_{ij}$ hypersurface $n_\mu=u_\mu$, so the derivations of Eq.~(\ref{eq: div and curl}) are done with respect to the fluid flow. 
Additionally, because of the nature of the Levi-Civita tensor and the symmetrisation applied to the curl, $(\nabla\times E)_{\mu\nu}$ only has spatial components. 
We compute the magnitude of these tensors following:
$|T| = \sqrt{g^{\alpha\mu}T_{\alpha}T_{\mu}}$ or $|T| = \sqrt{g^{\alpha\mu}g^{\beta\nu}T_{\alpha\beta}T_{\mu\nu}}$.

Fig.~(\ref{fig: 3d_EandB}) shows the $|E|$ and $|B|$ distribution in 3-D. 
These are made dimensionless by dividing by $H^2$.
The electric part is strongest along the vertices of the OD gradually moving towards the peak of the OD. 
To some extent, the electric part is analogous to the Newtonian description of gravity as it embodies tidal gravitational pull. 
The regions experiencing this the strongest are along the vertices as the matter is being pulled along the filaments towards the centre of the OD. 
At the peak, where the curvature is strongest, $|E|$ is small as the matter is already at the bottom of the potential well.

Conversely, the magnetic part is strongest around the vertices.
The filaments along the vertex direction, connecting the ODs periodically present, can be perceived, by analogy to electromagnetism, to be carrying a gravitational current, with $|E|$ strong along it, and $|B|$ strong around it.
In perturbation theory, the magnetic part is only constructed from vector and tensor modes and embodies relativistic effects. 
When we set the initial conditions, as explained in Section~\ref{sec: fully nonlinear initial conditions}, the density is defined non-linearly from the Hamiltonian constraint and the simulation freely evolves in full General Relativity. 
At nonlinear order the scalar, vector and tensor perturbations couple, explaining the non-zero magnetic part. 
Connecting this to the fluid flow, the magnetic part in general is sourced by shear, vorticity and acceleration \cite{G.F.R.Ellis_etal_2012}. 
Yet, in the synchronous-comoving gauge and with pressureless dust there is no vorticity or acceleration. 
Therefore, in this case, the magnetic part embodies the curl of the shear
\begin{equation}
    B_{\alpha\beta} = (\nabla\times \sigma)_{\alpha\beta},
\end{equation}
and we have shown the shear to be present in Fig.~(\ref{fig: Ray_slice}) and Fig.~(\ref{fig: Comoving_sphere}). 

On the leftmost panels of Fig.~(\ref{fig: 2dEB_div_curl}) the dimensionless $|E|$ and $|B|$ distributions are shown on a 2d plane, where the notable axes of symmetry are marked in the top panel. 
These are to be compared with Fig.~(\ref{fig: 3d_EandB}) to grasp these distributions. 
$|E|$ is indeed strongest along the OD vertex, black full line, and $|B|$ wraps around it. 
However we also see that they become negligible in the UD, and along the faces directions, dashed grey lines, and $|B|$ also disappears in the UD vertex direction, white full line. 
These axes of symmetry are notable features in the divergence and curl distributions, middle and right panels. 
The divergence is strongest close to the peak of the OD, and to the other OD present through periodic boundaries. 
Then the curl of $|B|$ is strongest along the vertex and the curl of $|E|$ is strongest around the vertex axis.

The presence of $|B|$ in itself is not proof of the benefit we get from having a fully relativistic simulation, as frame-dragging can be measured from Newtonian simulations \cite{I.Milillo_etal_2015, C.Rampf_etal_2016, M.Bruni_etal_2014_Feb, D.B.Thomas_etal_2015_16Jul} as well as in relativistic simulations \cite{J.Adamek_etal_2016_Mar, C.Barrera-Hinojosa_etal_2021_Jan, C.Barrera-Hinojosa_etal_2021_Dec}. 
However, when only gravitational waves are present $|E|=|B|$ \cite{W.B.Bonnor_1995}, the divergences vanish and the curls are present \cite{P.A.Hogan_G.F.R.Ellis_1997}. 
We look at Fig.~(\ref{fig: 1dEB_div_curl}) to see that here the domain average divergence does not vanish, and looking at the ratios, $|B|$ is smaller than $|E|$ but still has a per cent level presence. 
We also find that for the electric part, the domain average of the divergence is stronger than that of the curl, $\langle|\nabla\cdot E|\rangle_{\mathcal{D}}>\langle|\nabla\times E|\rangle_{\mathcal{D}}$, and the reverse is true for the magnetic part, $\langle|\nabla\cdot B|\rangle_{\mathcal{D}}<\langle|\nabla\times B|\rangle_{\mathcal{D}}$. 

The electric and magnetic parts of the Weyl tensor have previously been measured in numerical relativity cosmological simulations: \textit{i)} for a lattice of black holes, where the potential bias that is introduced by the magnetic part in optical measurements is quantified \cite{M.Korzynski_etal_2015, E.Bentivegna_etal_2018}; \textit{ii)} in more realistic cosmological simulations, where models with vanishing divergence of the magnetic part are found to be a valid approximation on large scales \cite{A.Heinesen_H.J.Macpherson_2022}. 
This differs from what we find as $\langle|\nabla\cdot B|\rangle_{\mathcal{D}}$ is initially present and grows throughout the simulation, even though it has the smallest amplitude in Fig.~(\ref{fig: 1dEB_div_curl}). 
These results do not directly contradict each other since we are considering very different spacial distributions, and here the simulation evolves into a highly nonlinear regime.

\subsection{Effective Petrov classification} \label{sec: Petrov classification}
\begin{figure}[th!]
    \centering
    \includegraphics[width=0.6\linewidth]{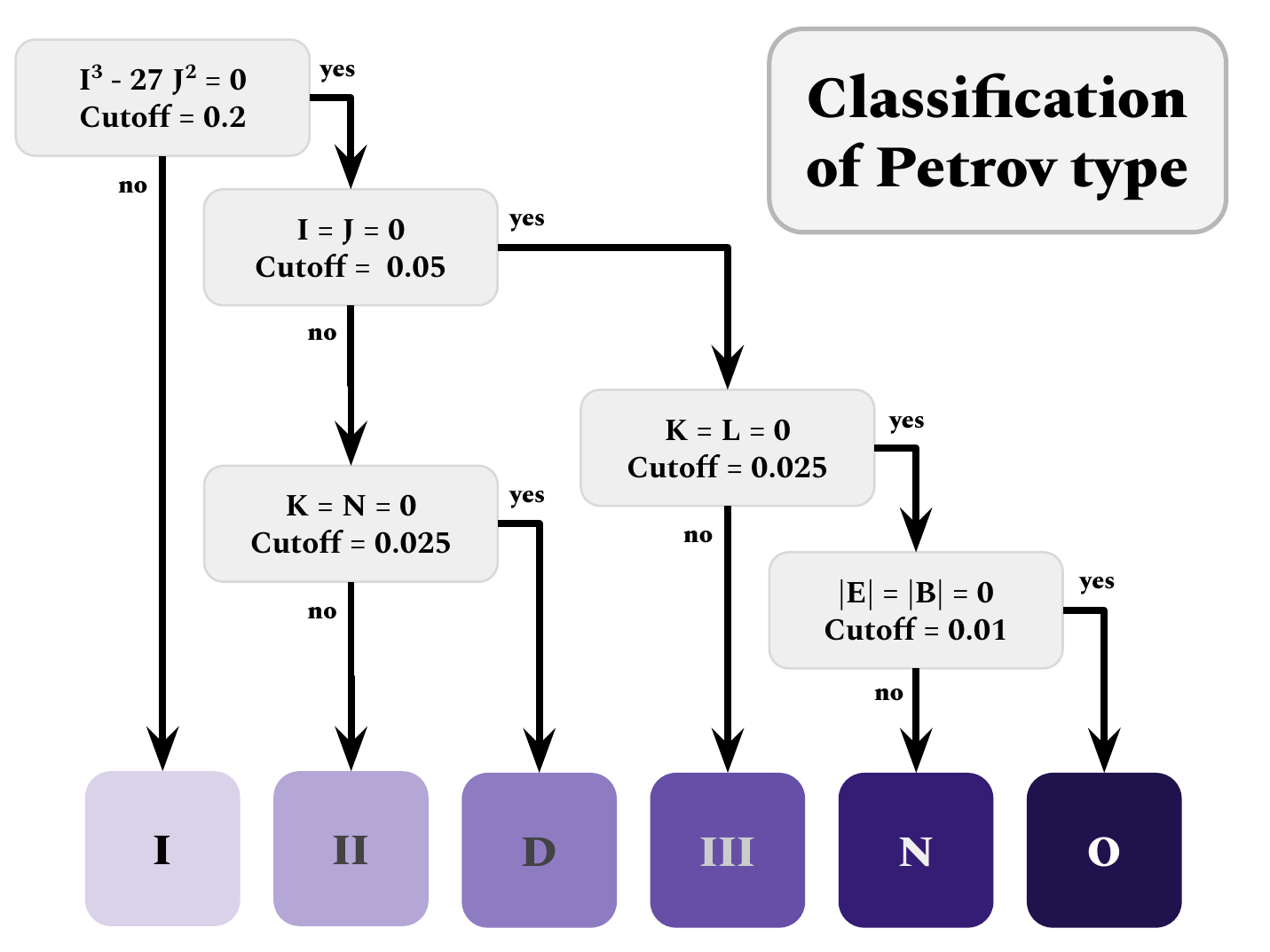}
    \caption{Flow diagram of Petrov classification, with a couple of modifications this is a replica of Fig.~(9.1) in \cite{H.Stephani_etal_2003}. Cutoff values used in our analysis are listed here.}
    \label{fig: petrovclassification}
\end{figure}

The Weyl tensor is the traceless part of the Riemann curvature tensor and describes, in essence, the tidal gravitational fields, far richer in a metric theory of gravity than in the Newtonian case. 
It is classified according to the Petrov classification \cite{A.Z.Petrov_2000}, with complex scalar invariants $I$, $J$, $K$, $L$, and $N$ that we compute from $E_{\alpha\beta}$ and $B_{\alpha\beta}$, following the methodology provided in Paper 1 \cite{R.L.Munoz_M.Bruni_2022, R.L.Munoz_2022_EBWeyl}. 
These invariants can then be used to classify different regions of the spacetime as Petrov type I, II, D, III, N, or O according to the scheme presented in Fig.~(\ref{fig: petrovclassification}), where we apply the theory of classification of exact solutions in \cite{H.Stephani_etal_2003}. 
Each Petrov type has a specific physical interpretation, e.g. type D is characteristic of the Schwarzschild and Kerr black holes, as well as of the tidal field outside a spherically symmetric gravitational field, while type N is characteristic of plane gravitational waves; we refer the reader to  Paper 1 \cite{R.L.Munoz_M.Bruni_2022, R.L.Munoz_2022_EBWeyl} and Refs.\ therein for more details. 

Numerically we hardly reach exact numbers, additionally, our simulation can be thought of as containing all types of perturbations at all orders, so our spacetime is of Petrov type I, the most general type. 
However, we consider the leading order type by introducing thresholds; then, because the background FLRW is of Petrov type O, that of conformally flat spacetimes, initially this is the leading order Petrov type, as the perturbations are initially small. 
As non-linearities grow, the spacetime becomes more general.
To see this transition, we adapt the \lstinline{IF} statements described in Fig.~(\ref{fig: petrovclassification}) by considering the real and imaginary parts of each quantity separately, normalising them, making them dimensionless, and comparing them to a chosen cutoff value. 
This is done by making these invariants have the same power as the Weyl tensor and dividing by $H^2$. 
For example for the real part of $I$, we then have the 

\begin{figure}[th!]
    \centering
    \includegraphics[width=0.99\linewidth]{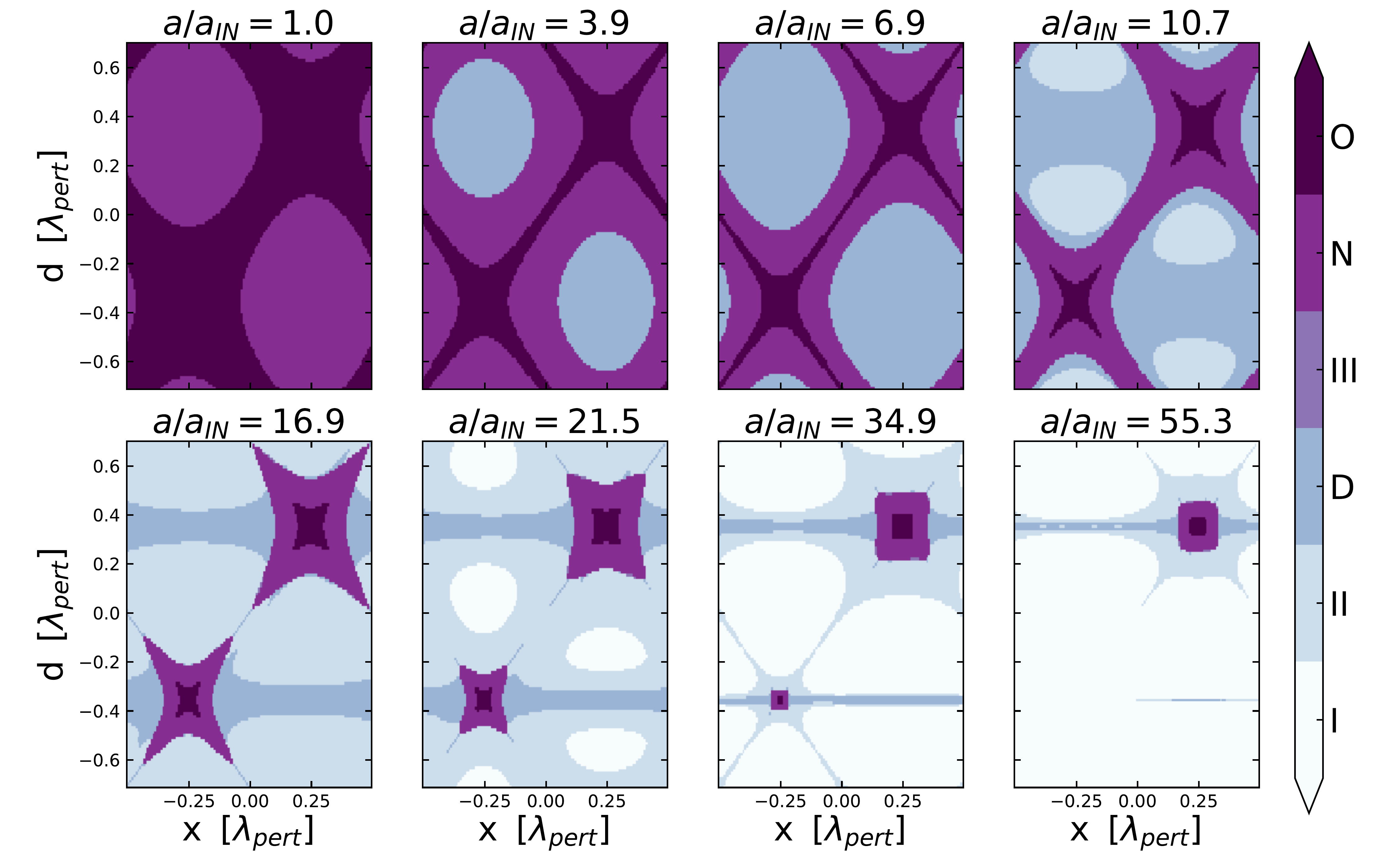}
    \caption{Classification of the spacetime regions according to the leading order Petrov type as defined in Fig.~(\ref{fig: petrovclassification}), along the $x$ and $y=z$ plane of the simulation box (with $d^2=y^2+z^2$). Eight points in time in the simulation are presented, and the corresponding normalised scale factor is on top of each panel. The peak of the over-density is in the bottom right quadrant, at $x=-0.25\lambda_{pert}$ and $d \simeq -0.35\lambda_{pert}$, it is periodically connected to other over-densities with a filament along the $d \simeq -0.35\lambda_{pert}$ direction, and the bottom of the under-density is in the top right quadrant at $x=0.25\lambda_{pert}$ and $d \simeq 0.35\lambda_{pert}$.}
    \label{fig: petrov}
\end{figure}

\noindent
value: $V = |Re(I^{1/6})|/H^{2}$, that we compare to a cutoff $V < c$. 
We also consider the numerical error obtained with the lower resolution simulations, see Appendix~\ref{sec: Constraints errors convergence}. 
So we adapt the statement to \lstinline{V < c  AND  ( V > Verror  OR  c > Verror)} where the part in parenthesis, establishes how reliable the variable is, if it isn't reliable we keep the classification general. 

The cutoff value is an arbitrary choice, if it is too small the whole spacetime is of type I, if it is large then it is of type O. 
No matter the choice of cutoff value the order of transition between the Petrov types remains the same, we then choose the cutoff values as presented in Fig.~(\ref{fig: petrovclassification}) to emphasise this behaviour. 
The cutoff values are not the same at all stages as we disentangle leading order contributions.

Following this process, Fig.~(\ref{fig: petrov}) shows the leading order spacetime on the $x$ and $y=z$ plane throughout the simulation. 
Overall the simulation starts as an effective type O spacetime, that of FLRW, as all the inhomogeneities embodied in the invariant scalars are all below the cutoff values; then the spacetime gradually transitions towards type I. 
This sort of peeling-off \cite{R.D'Inverno_1992, M.Alcubierre_2008} goes as O $\rightarrow$ N $\rightarrow$ D $\rightarrow$ II $\rightarrow$ I, from most special to least special. 
In this transition, we pass through all these spacetime types, with notable features related to the OD structure at hand.

Throughout this evolution, the peak of the OD and bottom of the UD are type O. 
These regions are conformally flat, which is not what we expected \textit{a priori} from the peak of the OD. 
However, as we saw previously, in this location $|E|=|B|=0$, therefore the spacetime is type O and the spatial curvature is non-zero, but the conformal curvature is constant as a local closed FLRW. 
Thus, this is another reason why the Top-Hat model describes the evolution of the peak of the OD very well.

Along the vertex direction, the transition goes as O $\rightarrow$ N $\rightarrow$ D. 
The focus of a D spacetime along the filament is interesting as this group includes the Schwarzschild, Kerr, and Szekeres metrics. 
The Weyl tensor of type D spacetimes has been described \cite{P.Szekeres_1965, R.D'Inverno_1992} as a Coulomb-like tidal field, where the matter gets elongated in a given direction towards a gravitational source, see Paper 1 \cite{R.L.Munoz_2022_EBWeyl} for more details. 
Indeed, we find that along the filaments matter is being pulled towards the two OD peaks they connect.

Then, remarkably, we note the strong presence of type N, the spacetime of gravitational waves. 
A non-spherically symmetric collapse is naturally expected to generate gravitational waves; here, we see tensor modes having a temporary leading order presence.
We leave the study of the generation of gravitational waves in nonlinear structure formation in full numerical relativity to future work.

\section{Conclusions} \label{sec: Conclusion}
In this work we have presented numerical relativity simulations of a simple nonlinear inhomogeneous structure growing in a $\Lambda$CDM universe. 
The simulations are run with the Einstein Toolkit \cite{F.Loffler_etal_2012, S.R.Brandt_etal_2020} using the new publicly available ICPertFLRW thorn \cite{R.L.Munoz_2023_ICPertFLRW}, then post-processed with our EBWeyl code \cite{R.L.Munoz_2022_EBWeyl} described in Paper 1 \cite{R.L.Munoz_M.Bruni_2022}.
We have used the synchronous-comoving gauge, i.e.\ the rest frame of CDM, represented as a pressureless and irrotational perfect fluid. 

The inhomogeneities are introduced with the comoving curvature perturbation $\mathcal{R}_c$, defined as a 3-D sinusoidal. 
This creates a periodic lattice of over-densities (OD) connected by filaments and surrounded by under-dense (UD) voids.
Near the peak of the OD the distribution of the matter and other fields is close to spherical symmetry, but this is no longer the case further out, as the structure tends towards an octahedron-like symmetry, with OD filaments along the vertices.

We obtain three main results: {\it i)} using $\mathcal{R}_c$, a gauge-invariant curvature perturbation typically used in early universe perturbation theory \cite{K.A.Malik_D.Wands_2008}, we successfully implement a purely growing mode in our initial conditions, following \cite{M.Bruni_etal_2014_Mar, M.Bruni_etal_2014_Sep}; in particular we use $\mathcal{R}_c$ to set up our initial metric and extrinsic curvature inhomogeneity, the fully nonlinear 3-Ricci curvature ${}^{(3)}R$, then defining the fully nonlinear matter density field from the Hamiltonian constraint, which is then automatically satisfied; {\it ii)} we study the evolution of the peaks through turn-around and collapse, finding that it is very well described by the Top-Hat model, to a level better than 1\%, see Table \ref{tab: Collapse}; {\it iii)} we study the Weyl tensor, both from the perspective of the electric and magnetic parts $E_{\alpha\beta}$ and $B_{\alpha\beta}$ and through a novel dynamical Petrov classification, finding that the gravito-magnetic effects are stronger around the filaments, and Petrov type N, the signature of gravitational waves, emerges in the directions connecting the OD peaks with the UD.

More in details, the main points are the following. 

\begin{itemize}
    \item The configuration described above leaves us free to choose the initial amplitude and wavelength of the inhomogeneities, as well as the initial redshift. 
    These are chosen such that initially we are in the linear regime and the simulation remains within the matter-dominated era (i.e.\ $\Lambda$ is negligible), even if our treatment is fully nonlinear. 
    Additionally, we identify the curvature-dominated regime, when the physical wavelength is larger than the Hubble scale, see Eq.~(\ref{eq: Long_wavelength}), a regime where the volume element is larger than the background in the OD region. 
    \item Monitoring the peak of the OD we find that, in this specific location, the turn-around (TA) and collapse are reached when the linearly extrapolated density contrast $\delta^{(1)}$ has values $\delta^{(1)}_{TA} = 1.05584 \pm 8\times 10^{-5}$ and $\delta^{(1)}_C = 1.676 \pm 2\times 10^{-3}$ in the  $\Lambda$CDM case, within 1\% of the   theoretically predicted values in the Top-Hat spherical and homogeneous collapse model \cite{V.Sahni_P.Coles_1995, P.Monaco_1998, J.A.Peacock_1999, H.Mo_etal_2010, N.Vittorio_2018}. 
    We explain this by looking at the contribution of the different terms in the Raychaudhuri equation, finding that the shear is in general subdominant around the peak and totally negligible at the peak, so that  at this location  the evolution is independent of its environment and in essence described by the Friedmann equations of a closed (positively curved) model. Indeed, our analysis also shows that at the peak location $\gamma^{1/3}{}^{(3)}R$ is constant in time, generalising into the fully nonlinear regime the conformal-curvature variable $\mathcal{R}_c$. 
    However, when considering a comoving sphere with a large comoving radius, containing a more significant inhomogeneity, its evolution can no longer be well described with the Top-Hat model.
    \item The peak of the OD is the first location to reach TA, when the expansion scalar reaches $\Theta=0$, then the surface $\Theta=0$ expands outward in the neighbouring region. 
    This TA boundary distinguishes an infalling and an expanding region. 
    The infalling region encompasses more and more material, eventually taking the shape of the entire OD region. 
    In the direction where $\delta$ is the biggest the TA radius increases the most, and in directions going through an UD region the TA radius eventually stops growing and shrinks instead. 
    These features are due to the inhomogeneous non-spherical shape we are working with. 
    \item Filaments are a fundamental part of the structure of the cosmic web, due to tidal fields \cite{J.R.Bond_etal_1995}. In computing the  electric and magnetic parts $E_{\alpha\beta}$ and $B_{\alpha\beta}$ of the Weyl tensor with EBWeyl \cite{R.L.Munoz_M.Bruni_2022, R.L.Munoz_2022_EBWeyl}, we find that $E_{\alpha\beta}$ is strongest along the filaments periodically connecting the ODs, stretching matter towards the OD centres, while $B_{\alpha\beta}$  wraps around the filaments. 
    On average the magnetic part is smaller than the electric part, with the ratio changing from $<10^{-2}$ to almost $10\%$ during the evolution.
    The divergence of $E_{\alpha\beta}$ is stronger than $E_{\alpha\beta}$ itself, while the curl of $B_{\alpha\beta}$ is stronger than $B_{\alpha\beta}$.
    For both, the divergence is strongest towards the OD, and the curl of $E_{\alpha\beta}$ is strongest on the filaments while the curl of $B_{\alpha\beta}$ is strongest around them.
    \item We also use EBWeyl \cite{R.L.Munoz_M.Bruni_2022, R.L.Munoz_2022_EBWeyl} to classify the spacetime as Petrov type I. However, introducing  a novel dynamical Petrov classification using thresholds that define leading order contribution, we find that the centres of the OD and UD are of type O, i.e.\ conformally flat as an FLRW model at leading order, while the spacetime is type D along the filaments, representing a simple tidal stretching along these directions, and transition as O $\rightarrow$ N $\rightarrow$ III $\rightarrow$ II $\rightarrow$ I elsewhere, with a notable presence of type N, typical of gravitational waves.
\end{itemize}
We believe that several interesting questions should be investigated as a follow-up to this work. 
Here we have neglected vorticity, for the good reason that it vanishes for purely scalar first-order perturbations while it is typically sourced in the multi-stream regime following the first shell crossing \cite{S.Pueblas_R.Scoccimaro_2009}, and it is a subdominant source for gravito-magnetic effects in N-body simulations \cite{M.Bruni_etal_2014_Feb, D.B.Thomas_etal_2015_16Jul, C.Barrera-Hinojosa_etal_2021_Jan}, also in $f(R)$ gravity \cite{D.B.Thomas_etal_2015_30Jul}. 
A rough test-field estimate suggests that even if vorticity were initially present at the peak of the OD, its value at the last reliable step of our simulations would only be about twice its initial value. 
However, it would be interesting to study the effect of vorticity in detail, cf.\ \cite{G.F.R.Ellis_etal_1989}, using a more general gauge.   
Considering that close to the OD peaks and around UD voids the spacetime is close to spherical symmetry, it would be interesting to extend our work to look for self-similar behaviour \cite{E.Bertschinger_1985_p1, E.Bertschinger_1985_p39, B.Jain_E.Bertschinger_1995}. 
Here we have considered an over-simplified structure based on a single initial wavelength: with this or starting from a more complex structure, the effects of different wavelengths, mode-coupling during nonlinear evolution \cite{B.Jain_E.Bertschinger_1993} and the effects of very large-scale tidal fields \cite{A.S.Schmidt_etal_2018} should be the subject of further investigations.

Finally, let us note two important points. 
First, in this paper we have confirmed how good the Top-Hat description of collapse is at the peak of the OD. 
We believe that this result is robust for profiles around the peak that tend to be spherically symmetric, but the analysis here should  be extended in two directions: to model the effects of different quasi-spherical profiles on virialisation \cite{D.Rubin_A.Loeb_2013}, and to understand the effects of introducing some anisotropy at the peak, in particular to measure how large the change of collapse time due to shear would be. 
Last but not least, is the issue of how to best set up initial conditions for large-scale structure simulations in order to optimise computational efficiency while maintaining the required accuracy of modelling in the era of precision cosmology. 
Historically many approximations have been introduced to model quasi-linear stages \cite{V.Sahni_P.Coles_1995, P.Monaco_1998}, and more recently to take into account relativistic effect \cite{C.Fidler_etal_2016, C.Fidler_etal_2017}. 
Various quasi-linear relativistic approximations have been considered in the past \cite{S.Matarrese_etal_1993, S.Matarrese_etal_1994_Jan, S.Matarrese_etal_1994_Mar, L.Kofman_D.Pogosian_1995, L.Hui_E.Bertschinger_1995, M.Bruni_etal_1995_Jun, M.Bruni_etal_1995_Jul, R.Maartens_etal_1996, C.F.Sopuerta_etal_1998} and more recently \cite{M.J.Pareja_M.A.H.MacCallum_2006, H.Y.Ip_F.Schmidt_2016, A.Heinesen_H.J.Macpherson_2022}; we believe that these should be further investigated, in order to understand how to improve the setting up of initial conditions for the modelling of relativistic effects in nonlinear stages of structure formation, cf. \cite{E.Quintana-Miranda_etal_2023}.

\begin{acknowledgments}
We thank Kazuya Koyama and Helvi Witek for useful suggestions during the development of this work. MB thanks Robert Brandenberger for a useful conversation about super-horizon modes and he is very grateful to Chul-Moon Yoo for pointing out the error in \cite{E.Bentivegna_M.Bruni_2016}, leading to misinterpretation, described in footnote 6 above.
We are also grateful to Hayley J. Macpherson, Jascha Schewtschenko, and Ian Hawke for useful discussion.
In this work MB has been supported by UK STFC Grant No. ST/S000550/1 and  ST/W001225/1, 
RLM has been supported by UK STFC studentship grants ST/S505651/1 and ST/T506345/1 and by University of Portsmouth funding.
Numerical computations were done on the Sciama High-Performance Compute (HPC) cluster \cite{Sciama} which is supported by the ICG, SEPNet and the University of Portsmouth.
For the purpose of open access, the author(s) has applied
a Creative Commons Attribution (CC BY) licence to any Author Accepted Manuscript version arising. Supporting
research data are available on reasonable request from the  authors.
\end{acknowledgments}

\bibliography{Ref}

\appendix
\section{Numerically Integrating} \label{sec: Num_int}
The average over a certain domain $\mathcal{D}$ of a scalar $\phi$ is computed as:
\begin{equation}
    \langle \phi \rangle_{\mathcal{D}}=\frac{\Delta x^3}{V}\sum_{\mathcal{D}} \phi \gamma^{1/2}
\end{equation}
with $\gamma$ the determinant of the spatial metric in our synchronous-comoving gauge and $\Delta x= \Delta y=\Delta z$ are the space coordinate intervals between grid points.
$V$ is the proper volume given by
\begin{equation}
    V = \Delta x^3\sum_{\mathcal{D}} \gamma^{1/2}.
\end{equation}
The proper and comoving lengths along a grid line are calculated by
\begin{equation}\label{eq:vertexlength}
    L_p=\Delta x\sum_{i=0}^{i_{max}} \gamma^{1/6}
    \;\;\;\;\;\;\text{and}\;\;\;\;\;\;
    L_c=\Delta x\sum_{i=0}^{i_{max}} 1 ,
\end{equation}
since $\Delta x$ is the comoving spatial coordinate element.
In the background the comoving length is related to the proper length simply by the scale factor: $L_p = a(t) L_c$. 

Computing $L_p$ and $L_c$ as in Eq.~(\ref{eq:vertexlength}) is perfectly fine along the vertex direction because this direction is aligned with the grid. 
However this is no longer the case in the face and edge directions, so a weighted integration is needed: 
\begin{equation}
    L_p = \Delta x \sum_{i=0}^{i_{max}} w \gamma^{1/6}
    \;\;\;\;\;\;\text{and}\;\;\;\;\;\;
    L_c = \Delta x \sum_{i=0}^{i_{max}} w
\end{equation}
with the weight $w$ in the range $0 \leqslant w \leqslant \sqrt{3}$. 
Each data point is in the centre of a cubic grid cell, so the value of this data point only applies to the section passing through this cell. 
$w \; \Delta x$ then represents the comoving length of the section contained in each cell. 
It is computed by finding the intersection between the integrated direction and the grid cells and then finding the length between these intersection points.

On occasion, we integrate up to $K=0$, or up to a given comoving radius; in these cases, the last weight to be used is measured between the last intersection and this boundary point. 
In both these cases, the boundary point is found using a trilinear interpolation within this last cell.

The chosen averaging domain in Section~\ref{sec: Comoving sphere}, is a comoving sphere. Approximating a sphere on a grid can be done by only considering the grid points contained within the sphere, however, we refine this with a weighted integration:
\begin{equation}
    \langle \phi \rangle_{\mathcal{D}}=\frac{\Delta x^3}{V}\sum_{\mathcal{D}} w \phi \gamma^{1/2}
    \;\;\;\;\;\;\text{and}\;\;\;\;\;\;
    V = \Delta x^3\sum_{\mathcal{D}} w \gamma^{1/2},
\end{equation}
with $0 \leqslant w \leqslant 1$. 
Here $w \; \Delta x^3$ is the comoving volume of the part of the cubic grid cell that is included in the comoving sphere. 
The weight $w$ is computed with the {\tt sphereint} code \cite{R.L.Munoz_2022_sphereint}, where the value of $w$ depends on the number of cubic grid cell vertices contained in the sphere, if all eight are in the sphere $w=1$, and if there are none $w=0$. 
When the cell is partially within the sphere, we compute the intersecting points, of the sphere and the cube edges, approximate the spherical boundary contained in the cube as a plane, and compute the volume of the corresponding geometry. 
Most cases take the form of trirectangular tetrahedrons. 
That is clear when one cube vertex is in the sphere, but in other cases, the shape is extended to be a trirectangular tetrahedron, and then smaller trirectangular tetrahedrons are removed. 
When four cube vertices are in the sphere there is a particular case where a truncated right square prism needs to be considered.

\section{Constraints, error bars and convergence}\label{sec: Constraints errors convergence}
The 3+1 decomposition of Einstein's field equations \cite{T.W.Baumgarte_S.L.Shapiro_2010} provide the Hamiltonian and momentum constraints:
\begin{equation}
    \mathcal{H} = {}^{(3)}R + \frac{2}{3}K^2 - 2A^2 - 2\Lambda - 2\kappa\rho = 0,
    \;\;\;\;\;\text{and}\;\;\;\;\;
    \mathcal{M}^{i} = D_j\left(A^{ij}-\frac{2}{3}\gamma^{ij}K\right)-\kappa J^i = 0,
\end{equation}
with $J^i=-\gamma^{ia}n^{b}T_{ab}$ is the momentum density, and $D_j$ the spatial covariant derivative.

\begin{figure}[t]
    \centering
    \includegraphics[width=\linewidth]{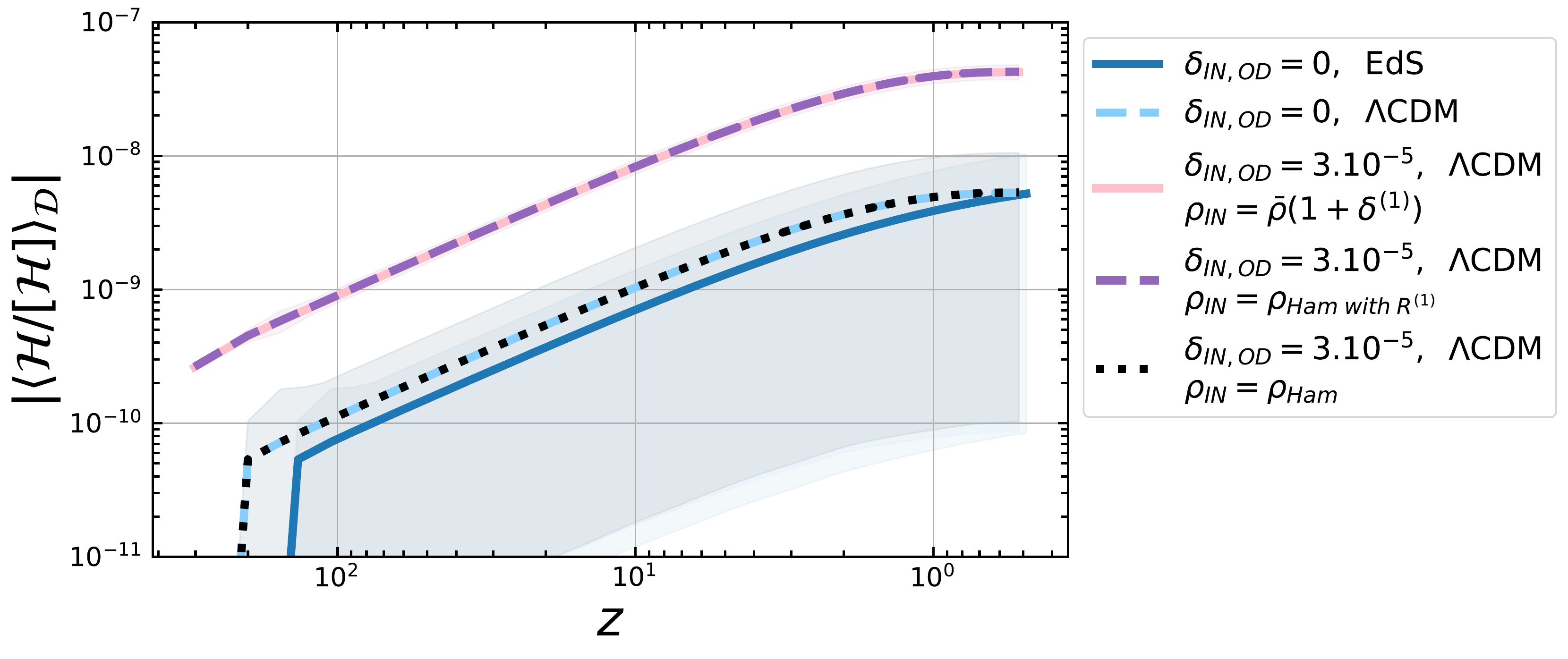}
    \caption{Domain average violation to the Hamiltonian constraint normalised with its energy scale of 5 different simulations, versus the redshift $z$. The initial (IN) amplitude of density contrast $\delta$ at the  peak of the over-density (OD) and the presence of $\Lambda$ in the simulations is specified in the legend. When $\delta_{IN, \;OD} = 3 \times 10^{-5}$, the initial energy density can be defined as $\rho_{IN} = \bar{\rho} (1 + \delta^{(1)})$ (\textit{pink full}), or as $\rho_{IN} = \rho_{Ham\; with \; R^{(1)}}$ (\textit{purple dashed}) from the Hamiltonian constraint but with first order 3-Ricci scalar, Eq.~(\ref{eq: Rc_comoving_curvature_perturbation}). We find that a better definition is $\rho_{IN}=\rho_{Ham}$ (\textit{black dotted}), in full from the Hamiltonian constraint using the first order $\gamma_{ij}$ and $K_{ij}$, Eq.~(\ref{eq: metricRc}) and Eq.~(\ref{eq: curvRc}), but the fully nonlinear 3-Ricci scalar of $\gamma_{ij}$. Here $\lambda_{pert}=1821$Mpc and $z_{IN}=302.5$ for $\Lambda$CDM initially and $\lambda_{pert}=1206$Mpc and $z_{IN}=205.4$ otherwise. Error bars, when visible, are indicated as shaded regions.}
    \label{fig: Constraint}
\end{figure}

We estimate the accuracy of the initial conditions implemented by quantifying the violation of these constraints ($\mathcal{H}$, or $\mathcal{M}^{i}$) normalised with their relative energy scales \cite{J.B.Mertens_etal_2016, H.J.Macpherson_2019}:
\begin{equation}
    [\mathcal{H}] = \left[ \left({}^{(3)}R\right)^2 + \left(\frac{2}{3}K^2\right)^2 + \left(2A^2\right)^2 + \left(2\Lambda\right)^2 + \left(2\kappa\rho\right)^2 \right]^{1/2},
\end{equation}
\begin{equation}
    [\mathcal{M}] = \left[ D_j(A^{ij})D_j(A_{i}^{j})+\left(\frac{-2}{3}\right)^2\gamma^{ij}D_j(K)D_i(K)+(-\kappa)^2 J^i J_i \right]^{1/2}.
\end{equation}
The momentum constraint is automatically satisfied at first order so we first focus on the Hamiltonian constraint as presented in Fig.~(\ref{fig: Constraint}). 
This enables us to try different methods to set the initial conditions of the simulation and find the best approach.

Firstly, we consider pure FLRW simulations ($A_{pert}=0$) in both the $\Lambda$CDM Eq.~(\ref{eq: LCDM}) and EdS Eq.~(\ref{eq: EdS}) models. Their normalised $\mathcal{H}$, domain averaged over the whole simulation box, are presented as blue lines in Fig.~(\ref{fig: Constraint}). 
In both cases, we find a small error confirming these were implemented correctly. 

Secondly, various methods of implementing the perturbation in the energy density are tried. 
We show the impact of initially setting $\rho$ up to it's first order as $\rho_{IN}=\bar{\rho}(1+\delta^{(1)})$ using Eq.~(\ref{eq: delta1}), this is the pink curve. 
Then we show the impact of including higher order terms by defining $\rho$ with the Hamiltonian constraint Eq.~(\ref{eq: rhoHam}), this is the dotted black line. 
Where all terms on the right-hand side of Eq.~(\ref{eq: rhoHam}) are calculated in full from the definition of $\gamma_{ij}$ and $K_{ij}$, Eq.~(\ref{eq: metricRc}) and Eq.~(\ref{eq: curvRc}). 
This shows a significant decrease in the initial error, for this perturbation amplitude, it even matches the simulations without perturbations. 

We highlight the importance of including the higher order terms consistently, with the purple dashed line, where $\rho$ is initially defined from the Hamiltonian constraint but instead of being calculated in full from the metric, the 3-Ricci is provided using only the first order expression, Eq.~(\ref{eq: Rc_comoving_curvature_perturbation}). The error in the resulting simulation matches that of the simulation with only first-order terms. So the best approach, that we use for our simulations, corresponds to the dotted black line with $\rho$ obtained from the Hamiltonian constraint in full.

The error bars on Fig.~(\ref{fig: Constraint}), and throughout, are obtained by using two other simulations of double grid size each, such that we have 3 simulations, each of $32^3$, $64^3$, and $128^3$ data points. Consider the result $f_{\Delta x}$ from a simulation with grid size $\Delta x$, we have accompanying simulations of grid size $2\Delta x$ and $4\Delta x$ each having their respective solution $f_{2\Delta x}$ and $f_{4\Delta x}$. The error on $f_{\Delta x}$ is then \cite{M.Alcubierre_2008}:

\begin{figure}[t]
    \centering
    \includegraphics[width=\linewidth]{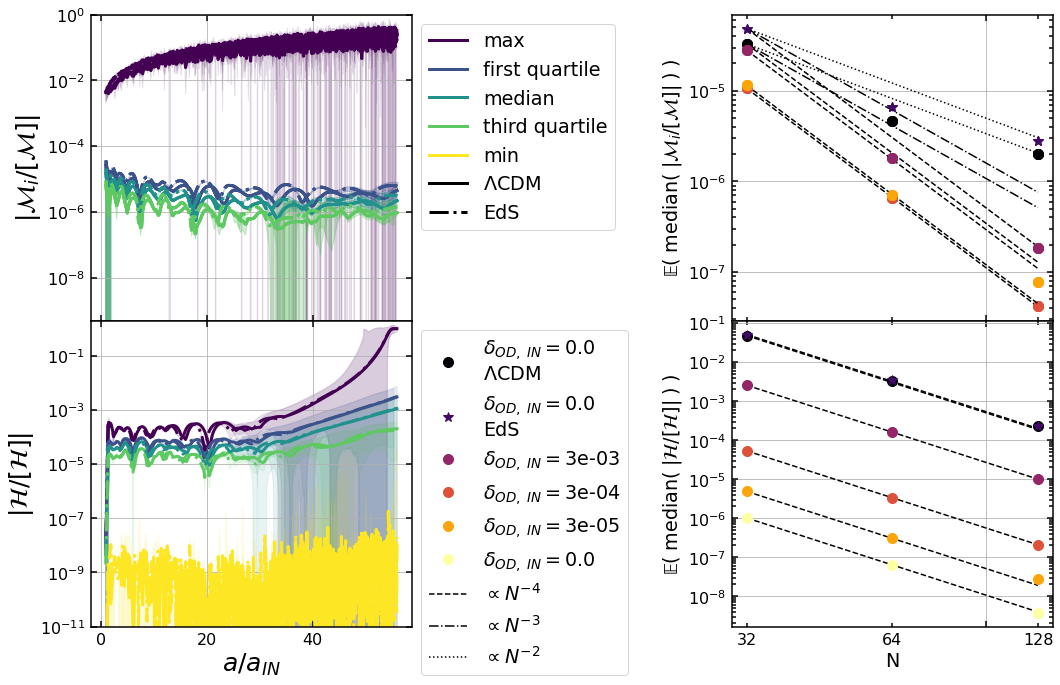}
    \caption{\textit{Left:} momentum (top) and Hamiltonian (bottom) constraint violation normalised with their respective energy scale measured at different quartiles of the data distribution during the evolution of the simulation. 
    The simulation with $\Lambda$ is indicated with full lines while the one without is indicated with dot-dashed lines. The three momentum constraints $i=\{1,\;2,\;3\}$ are all plotted with the same lines but all overlap so are not distinguishable. Error bars, when visible, are indicated as shaded regions.
    \textit{Right:} average median of these constraints for simulations of different resolution ($N^3$ the number of data points) and amplitude of the initial (IN) density contrast at the peak of the over-density (OD) $\delta_{IN, \; OD}$.
    When perturbed, the energy density is initially defined in full from the Hamiltonian constraint. $\lambda_{pert}=1821$Mpc and $z_{IN}=302.5$ when $\Lambda\neq0$ initially and $\lambda_{pert}=1206$Mpc and $z_{IN}=205.4$ otherwise.}
    \label{fig: Convergence}
\end{figure}

\begin{equation} \label{eq: Rerror}
    \epsilon_{\Delta x}=\frac{f_{2\Delta x}-f_{\Delta x}}{C-1}
\end{equation}
with the convergence
\begin{equation} \label{eq: convergence}
    C = \frac{|f_{4\Delta x}-f_{2\Delta x}|}{|f_{2\Delta x}-f_{\Delta x}|} = 2^n
\end{equation}
and $n$ is the order of the finite differencing approximation. 
$4^{th}$ order schemes are used for the simulation evolution and in post-processing, see Paper 1 \cite{R.L.Munoz_M.Bruni_2022, R.L.Munoz_2022_EBWeyl}. 

To check convergence in the simulations we show in Fig.~(\ref{fig: Convergence}) the error in the normalised Hamiltonian and momentum constraints. 
On the left panels, we plot their absolute value at different quartiles of the grid distribution, and then on the right, the average median is considered versus the resolution \cite{H.J.Macpherson_etal_2017}. 
The truncation error that comes from the finite difference schemes will fit a line, that is $\propto N^{-n}$, indicative of the convergence. 

For the Hamiltonian constraint, while the amplitude of the violation may increase as the perturbation amplitude increases, it still continues to follow $4^{th}$ order convergence, as expected.

For the momentum constraint, while the same could be said for small perturbations, the top right panel of Fig.~(\ref{fig: Convergence}) shows a decreased convergence when $\delta_{IN, \; OD} = 0.03$. 
Indeed the momentum constraint is only satisfied a first order, so in a nonlinear scenario, the solution tends towards a non-zero solution. 
However, the top left panel shows that while there is a violation of the momentum constraint, this does not grow throughout the simulation. 
The max curve may seem concerning but this is because it is amplified by data points whose momentum energy scale is the numerical equivalent of zero, thus the shape of the curve resembles numerical noise. 
In computing $C$, with Eq.~(\ref{eq: convergence}) we find the average convergence of the median normalised violation to the momentum constraint to be $C\simeq 13.76$ for the case with $\Lambda$ and $C\simeq 15.47$ for the case without, indicating that this solution has a $3.7 - 3.9$ order convergence towards a non-zero solution that does not grow during the simulation.

\end{document}